\documentclass[lettersize,journal]{IEEEtran}
\usepackage[hidelinks]{hyperref}
\usepackage{cite}
\usepackage{amsmath,amssymb,amsfonts}
\usepackage{algorithmic}
 \usepackage[x11names,dvipsnames,table,xcdraw]{xcolor} 
\usepackage{amssymb}
\usepackage{graphicx}
\usepackage{threeparttable}
\usepackage{textcomp}
\usepackage{pdfx}
\usepackage[normalem]{ulem}  
\usepackage{soul}
\usepackage[utf8]{inputenc}
\usepackage[linesnumbered,ruled,vlined]{algorithm2e}
\SetKwFunction{KwFunction}{\textbf{Function}}
\usepackage{amsthm}

 \usepackage[framemethod=tikz]{mdframed}
 \usepackage{booktabs}
 \usepackage{makecell}
 \usepackage{colortbl}
 \usepackage{pgf}

\usepackage{float} 
\usepackage{listings}
\usepackage{graphicx}
\usepackage{caption} 
\usepackage{subcaption} 
\usepackage{framed}
\usepackage{bm}
\usepackage{array}
\usepackage{multicol} 

\usepackage{makecell}
\usepackage{pifont}
\usepackage{enumerate}
\usepackage{multirow}
\usepackage{booktabs}
\usepackage{afterpage}
\usepackage{url}

\usepackage{etoolbox}

\usepackage{tcolorbox}

\newcommand{\redcheck}{\textcolor{red}{\scalebox{1.5}{$\checkmark$}}}

\usepackage{inconsolata}
\definecolor{codebg}{rgb}{0.95,0.95,0.95}
\definecolor{ForestGreen}{RGB}{34,139,34}

\newcommand{\method}{S$^2$F}

\begin{document}

\title{\method: Principled Hybrid Testing With Fuzzing, Symbolic Execution, and Sampling}

\author{Lianjing~Wang, Yufeng~Zhang, Kenli~Li, Zhenbang~Chen, Xu~Zhou, Pengfei~Wang, Guangning~Song, and Ji~Wang%
\thanks{L. Wang, Y. Zhang, and K. Li are with the College of Computer Science and Electronic Engineering, Hunan University, China (e-mail: lianjing@hnu.edu.cn; yufengzhang@hnu.edu.cn; lkl@hnu.edu.cn). (Corresponding author: Yufeng Zhang.)}%
\thanks{Z. Chen, X. Zhou, P. Wang, G. Song, and J. Wang are with National University of Defense Technology, China (e-mails: zbchen@nudt.edu.cn; zhouxu@nudt.edu.cn; pfwang@nudt.edu.cn; songguangning@nudt.edu.cn; wj@nudt.edu.cn).}%

}


\markboth{Journal of \LaTeX\ Class Files,~Vol.~14, No.~8, August~2021}%
{Shell \MakeLowercase{\textit{et al.}}: A Sample Article Using IEEEtran.cls for IEEE Journals}

\IEEEpubid{0000--0000/00\$00.00~\copyright~2021 IEEE}

\maketitle

\begin{abstract}
Hybrid testing that integrates fuzzing, symbolic execution, and sampling has demonstrated superior testing efficiency compared to individual techniques.	However, the state-of-the-art (SOTA) hybrid testing tools do not fully exploit the capabilities of symbolic execution and sampling in two key aspects. First, the SOTA hybrid testing tools employ tailored symbolic execution engines that tend to over-prune branches, leading to considerable time wasted waiting for seeds from the fuzzer and missing opportunities to discover crashes. Second, existing methods do not apply sampling to the appropriate branches and therefore cannot utilize the full capability of sampling.
To address these two limitations, we propose a novel hybrid testing architecture that combines the precision of conventional symbolic execution with the scalability of tailored symbolic execution engines. Based on this architecture, we propose several principles for combining fuzzing, symbolic execution, and sampling. 
We implement our method in a hybrid testing tool \method. To evaluate its effectiveness, we conduct extensive experiments on 15 real-world programs. Experimental results demonstrate that \method\ outperforms the SOTA tool, achieving an average improvement of 6.14\% in edge coverage and 32.6\% in discovered crashes. Notably, our tool uncovers three previously unknown crashes in real-world programs.
\end{abstract}


\maketitle
\section{Introduction}

Fuzzing and symbolic execution are two of the most widely applied techniques for bug detection and vulnerability discovery in both academia and industry \cite{ossfuzz, AFL, SAGE, KLEE}. Coverage-guided fuzzing \cite{AFL, fuzzroadmap, ossfuzz} selects a seed from a pool and generates multiple mutated inputs. The program under test is instrumented to identify inputs that increase code coverage; such inputs are then added to the seed pool for future mutations. 
Symbolic execution, in contrast, symbolically executes programs and collects path conditions for each explored branch \cite{symbolic_execution_survey, symbolic_execution_three, king, KLEE, SAGE, SPF, CUTE, S2E}. These path conditions are solved using constraint (or SMT) solvers \cite{STP_Solver, Z3}, and the solutions can guide execution toward specific program paths. Symbolic execution is particularly valued for its ability to explore deep program states governed by complex, nested conditions. Since CUTE\cite{CUTE} and DART \cite{DART}, researchers have combined concrete execution with symbolic execution to improve scalability and feasibility. This approach is called dynamic symbolic execution or concolic execution.

The philosophies of fuzzing and symbolic execution are inherently complementary.
To leverage the strengths of both, researchers have proposed integrating them into hybrid testing \cite{Driller, QSYM, SymCC, SymSAN, HybridFT, T-Fuzz, Dowser, Triton}.
The core idea behind hybrid testing is that \textit{program regions governed by complex conditions should be explored using costly symbolic execution, while easily reachable paths can be efficiently handled by fuzzing}.
Typically, a hybrid testing system employs a coordinator to manage the interaction between fuzzing and symbolic execution, enabling bidirectional seed exchange between the two components. When the coordinator selects a seed from the fuzzer, the symbolic executor performs symbolic execution with it and solves path constraints to generate new inputs. These newly generated seeds are then transmitted to the fuzzer for further mutation.
Empirical results consistently demonstrate that hybrid testing achieves higher testing efficiency than fuzzing or symbolic execution alone \cite{Driller, QSYM, Pangolin, Co_Fuzz}.

The development of hybrid testing has significantly influenced the architecture of symbolic execution engines.
Early hybrid systems combine coverage-guided fuzzing with conventional symbolic executors. Researchers soon realized that several key design features of conventional symbolic executors were not well-suited for hybrid testing. For example, conventional symbolic execution uses snapshots at each branch to avoid re-execution when exploring neighboring paths, but these snapshots cannot be reused in hybrid testing. Additionally, symbolic executors are often overwhelmed by the large number of seeds generated by fuzzers. To address this issue, Yun \textit{et al.} proposed QSYM \cite{QSYM}, which integrates a symbolic execution engine tailored for hybrid testing (hereafter referred to as \textit{tailored symbolic executor} in this paper\footnote{Although no consensus definition exists, several recent hybrid testing tools, including QSYM, SymCC, SymQEMU, SYMSAN, PANGOLIN, CoFuzz, \textit{etc.}, adopt the similar strategies in symbolic execution engine, which we refer to in this paper as a \textit{tailored symbolic executor}. }). QSYM eliminates several heavyweight mechanisms of conventional symbolic executors, such as execution tree maintenance and snapshots, and instead operates as a single-path symbolic executor invoked iteratively for each seed by the coordinator. Importantly, QSYM employs an excessive pruning strategy (discussed later) to improve testing efficiency.
Since then, modern symbolic executors in hybrid frameworks (\textit{e.g.}, SymCC \cite{SymCC}, SYMSAN \cite{SymSAN}, SymQEMU \cite{SymQemu}) adopt a similar strategy for efficiency.

\IEEEpubidadjcol

In parallel, several recently proposed hybrid testing tools, such as PANGOLIN \cite{Pangolin} and CoFuzz \cite{Co_Fuzz}, integrate sampling techniques with fuzzing to enhance testing efficiency. This approach samples multiple solutions from the solution space of a path condition at each branch \cite{Legion, Pangolin, Co_Fuzz}. The sampled inputs share the same path prefix but diverge in their path suffixes, guiding exploration toward targeted program regions. Consequently, sampling is often referred to as constrained fuzzing. While sampling at a branch is generally more costly than constraint solving, it produces multiple inputs instead of a single solution, thus making the average cost per input between fuzzing and symbolic execution. Additionally, sampling is more controllable than fuzzing but more stochastic than symbolic execution. In essence, sampling strikes a balance between the randomness of fuzzing and the precision of symbolic execution.

In this paper, we focus on enhancing the testing efficiency of hybrid testing by integrating fuzzing, symbolic execution, and sampling. 
We observe that the state-of-the-art (SOTA) hybrid testing tools fail to fully leverage the potential of symbolic execution and sampling, due to limitations in their fundamental architecture and coordination mechanism.  We improve the testing efficiency of hybrid testing by addressing the following two issues.

The first issue is the \textit{sleeping phenomenon} of symbolic executors in hybrid testing. 
Hybrid testing tools run fuzzers and symbolic executors in parallel. Ideally, each component should continuously explore program paths. However, we observe that symbolic executors in SOTA hybrid testing tools may waste a large portion of their running time. For example,
our experimental results show that the symbolic executor in the  SOTA hybrid testing tool, SymCC \cite{SymCC}, wastes 56.42\% of the total running time on waiting for seeds from the fuzzer. The main reason is tailored symbolic execution engine only maintains branch coverage rather than a global execution tree for efficiency. According to branch coverage, the pruning strategy over-prunes branches during exploration. Importantly, this pruning strategy is induced by the underlying architecture of the symbolic execution engine. Therefore, \textit{designing an architecture for symbolic execution in hybrid testing that can utilize the time resource effectively is still a challenge}.

To address the \textit{sleeping issue}, we propose a novel hybrid testing architecture. The architecture consists of two core components: (1) a coordinator that maintains a lightweight execution tree to track path coverage continuously, and (2) a tailored symbolic execution engine that explores each path efficiently. 
Through this design, the system can explore the branches that are excessively pruned by tailored symbolic executors in SOTA hybrid testing tools, thereby effectively mitigating the \textit{sleeping issue}. 
Meanwhile, the architecture supports flexible search strategies like conventional symbolic executors due to its maintained execution tree. In contrast, tailored symbolic executors cannot distinguish the path prefix of branches and are hence limited in search strategies.
Furthermore, our architecture keeps the scalability of tailored symbolic execution engines.

The second issue addressed in this paper concerns the principled integration of sampling into hybrid testing. Existing hybrid testing frameworks that employ sampling, such as PANGOLIN \cite{Pangolin} and CoFuzz \cite{Co_Fuzz}, replace constraint solving with sampling during symbolic execution. In symbolic execution, collecting and solving path conditions incur substantial overhead. Therefore, these two works employ sampling rather than a single constraint solving in order to exploit each collected path condition more thoroughly.
Nevertheless, we observe that while complex branches should indeed be explored via symbolic execution, not all branches justify the high cost of sampling.
Therefore, \textit{the principled and effective integration of sampling into hybrid testing remains a key challenge}.

To address this issue, we propose several principles of hybrid testing that integrate fuzzing, symbolic execution, and sampling.
Based on our proposed new architecture, we design an advanced coordination mechanism that implements these hybrid testing principles. Experimental results validate the rationality and effectiveness of the proposed integration strategy.

The main contributions of this paper are as follows.
\begin{enumerate}[(1)]

\item
We systematically evaluate SOTA hybrid testing tools and reveal that their symbolic execution engines may spend more than half of the total runtime waiting for seeds from the fuzzer. Through in-depth analysis, we identify that this inefficiency originates from the pruning strategy in tailored symbolic execution engines.

\item We propose a novel hybrid testing architecture, combining the flexibility of conventional symbolic executors and the scalability of tailored symbolic executors. The new architecture reduces the sleeping time of symbolic execution in hybrid testing significantly.

\item We establish a set of principles for integrating fuzzing, symbolic execution, and sampling, addressing the ``where to sample'' issue. These principles are instantiated in the coordination mechanism. 

\item We implement the proposed architecture and coordination mechanism in a new hybrid testing tool, \method. Extensive experiments on 15 real-world show that \method\ outperforms SOTA hybrid testing tool by 6.14\% in edge coverage and 32.6\% in discovered crashes.
\end{enumerate}

The remaining parts of this paper are organized as follows. Section \ref{sec:problem} discusses background and motivation.  Section \ref{sec:method} illustrates the details of our method.
Section \ref{sec:ImplEval} presents implementation details and experimental results. Section~\ref{sec:Discussion} presents discussion and future work.
Section \ref{sec:relatedwork} reviews related work. Finally, Section \ref{sec:conclude} concludes the paper.

\section{Background and  Motivations }\label{sec:problem}

\subsection{Background}

\begin{figure*}[ht!]
	\centering
	\includegraphics[width=1\linewidth]{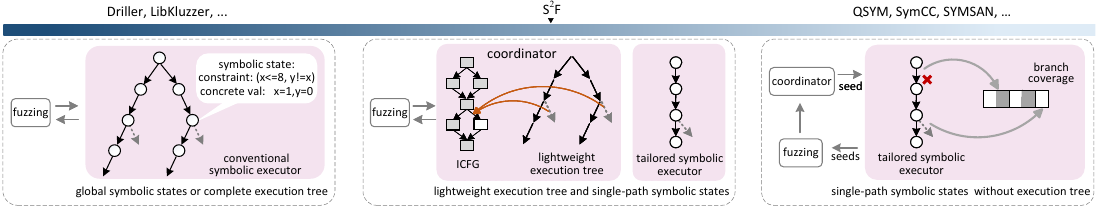}
	\caption{Spectrum of hybrid testing architectures.}
	\label{fig:spectrum}
\end{figure*}
\afterpage{\vspace*{-1pt}}

\textbf{Hybrid Testing.}
Coverage-guided fuzzing and symbolic execution are two of the most widely used software testing techniques. Coverage-guided fuzzing executes the program with mutated inputs and uses code-coverage feedback to guide the search toward previously unexplored execution paths. Symbolic execution executes programs symbolically and solves path condition with a constraint solver to generate new inputs.

Hybrid testing integrates these two complementary techniques by applying symbolic execution to handle branches that are difficult for fuzzing to cover. The central challenge in hybrid testing is effectively coordinating fuzzing and symbolic execution. Several approaches have been proposed to address this challenge. For example, DigFuzz~\cite{digfuzz_github} sends the hardest paths to symbolic execution. MEUZZ~\cite{MEUZZ} employs a regression model to predict seed utility values and guides seed scheduling. Tools such as SymCC~\cite{SymCC}, QSYM~\cite{QSYM}, and SYMSAN~\cite{SymSAN} select seeds for symbolic execution based on several seed features such as seed length, execution time, \textit{etc}.

\textbf{Sampling.}
Consider a branch $br$ with a strict path condition $\phi$ that dominates a substantial portion of the program. When symbolic execution solves $br$ and generates a seed $s$, the seed is then passed to the fuzzer for further mutation. However, random mutations applied to $s$ may easily violate the strict path condition $\phi$, preventing the resulting mutants from reaching the program parts below $br$.

To address this issue, researchers have introduced sampling into hybrid testing \cite{Pangolin, Co_Fuzz}.
During symbolic execution, sampling at a branch $br$ produces multiple inputs that satisfy the same path condition but diverge in their path suffixes. These sampled inputs can successfully explore the program regions dominated by $br$. 
Theoretically, sampling is more computationally expensive than constraint solving.

\subsection{Motivation 1:  The Sleep Issue of Symbolic Executor}\label{sec:moti1}

In this subsection, we first discuss the evolution of architectures of symbolic executors in hybrid testing. Next, we report the issue of SOTA hybrid testing tools, \textit{i.e.}, the symbolic executor may spend considerable time waiting for seeds from fuzzing. Finally, we point out that the underlying reason is the inadequate architecture of current hybrid testing tools.

\subsubsection[]{\textbf{Architectures of Symbolic Executor in Hybrid Testing}}
\textbf{Conventional Approach.} At its birth \cite{Driller}, hybrid testing integrates a coverage-guided fuzzer and a conventional symbolic executor in a straightforward way (the left part of Fig. 1). This architecture integrates a symbolic executor that maintains an execution tree or symbolic states of all paths, supporting all features of traditional symbolic execution. However, certain design elements of conventional symbolic executors are not well-suited for hybrid testing \cite{QSYM}. For example, Driller uses snapshots to reduce the overhead of re-executing a target program when exploring neighboring paths. In hybrid testing, the symbolic executor is invoked iteratively to re-execute seeds generated from fuzzing, and snapshots from different paths cannot be reused due to different variable assignments across paths.

\begin{table*}
	\centering
	\scriptsize
	\caption{\scriptsize Architectures, pruning/search strategies, and sampling strategy of hybrid testing tools. F: fuzzer. SE: symbolic executor. LOB: local opposite branch strategy (see Section \ref{sec:moti1}).  SOLVE: constraint solving.}
	\label{tab:existing_architecture_search_strategies}     
	\begin{tabular}{lcccc}
		\toprule
		Tool & Architecture &Pruning Strategy  & Search Strategy & Solve or Sample During Symbolic Execution \\
		\midrule
		Driller \cite{Driller}   & F + conventional SE   & from Angr              & from Angr  &all SOLVE \\ 
		libKluzzer \cite{LibKluzzer} & F + conventional SE        &from KLEE    &from KLEE      & all SOLVE     \\
		QSYM \cite{QSYM}       & F + tailored SE &LOB     & by simple seed features &all SOLVE   \\
		DigFuzz \cite{Dig_Fuzz}    & F + tailored SE &LOB    &     hard branches first                   &all SOLVE   \\
		MEUZZ \cite{MEUZZ}      & F + tailored SE &LOB  &  linear model on seed features &all SOLVE  \\
		SymCC \cite{SymCC}      &F + tailored SE &LOB     & by simple seed features & all SOLVE   \\
		SYMSAN \cite{SymSAN}     & F + tailored SE &LOB    & by simple seed features  & all SOLVE   \\
		PANGOLIN \cite{Pangolin}   & F + tailored SE &LOB    &  hard branches first  & all sampling  \\ 
		CoFuzz \cite{Co_Fuzz}     & F + tailored SE &LOB &linear model on simple branch features                  &  all sampling   \\
		\hline
		\textbf{\method}\ (our) 
        &\makecell{F + \textbf{lightweight execution tree} \\ \textbf{+ tailored SE}}
        & no pruning 
        &hard and valuable branches first  
        & \makecell{\textbf{sample on hard and valuable branches}, \\ \textbf{{solve rest branches}}} \\
		\bottomrule
	\end{tabular}
	\vspace{1mm}
	\captionsetup{font=footnotesize}

\end{table*}

\textbf{Tailored Approach.}
To overcome the limitations of conventional symbolic executors in hybrid testing, Yun \textit{et al.} use a symbolic executor tailored for hybrid testing in QSYM \cite{QSYM}. QSYM employs a coordinator to synchronize seeds between the fuzzer and the symbolic executor. When the coordinator selects an input from the seed pool of AFL, it starts a new symbolic execution process to execute a single path concolically. Instead of taking snapshots at each branch statement, the symbolic executor solves the unexplored opposite branch (referred to as \textit{open branch} or \textit{off-the-path branch} in the literature) to generate new input. 
Tailored symbolic executors also do not maintain an execution tree or global symbolic states. Instead, it only tracks branch coverage rather than path coverage.
Upon reaching the end of the path, the solved results at open branches are synchronized back to the fuzzer. Then the control flow returns to the coordinator. If all seeds from the fuzzer are processed, the coordinator waits for the fuzzer to generate new seeds. 
More features of the tailored symbolic executor can be referred to \cite{QSYM}.
In summary, the tailored symbolic executor adopts a design philosophy more suitable for hybrid testing and lies at the opposite end of the spectrum (see the right side of  Fig. \ref{fig:spectrum}).

Table \ref{tab:existing_architecture_search_strategies} compares the architecture and several other aspects of recently proposed hybrid testing tools, including QSYM, DigFuzz \cite{Dig_Fuzz}, SymCC \cite{SymCC}, MEUZZ \cite{MEUZZ},  CoFuzz \cite{Co_Fuzz},  SYMSAN \cite{SymSAN}, \textit{etc}.  These tools all inherit the design of the tailored symbolic executor, although they differ in other aspects.

\textbf{Pruning Strategy of SOTA Hybrid Testing Tools}.
As the first tailored symbolic executor, QSYM introduces a special pruning strategy to determine whether a branch should be solved when executing a seed. 
QSYM uses a bitmap shared across seeds to track branch coverage, where each branch is mapped to an index. During symbolic execution, if a branch $br$ on the current path brings new coverage (mode) \footnote{New coverage mode means the coverage count reaches a new value in its base-2 logarithm. See AFL \cite{AFL} for details.}, its opposite branch $br\prime$ is solved to generate new input. Then both $br$ and $br\prime$ are marked as covered in the bitmap. Otherwise, $br\prime$ is pruned. We refer to such pruning strategy as \textit{\underline{Local Opposite Branch (LOB) strategy}}\label{lin:LOB}, as it decides whether to prune an open branch according to the local coverage of its opposite branch \footnote{Although SymCC/QSYM also uses a separate \textit{bitmap} to store the context of each branch. But the stored context is limited to the latest two branches on the current path. Furthermore, the context bitmap is filled quickly in practice. So the context information is limited and cannot act as path coverage.}.
 
The LOB strategy may prune branches that would otherwise be explored in conventional symbolic executors.
Consider that a branch $br$ is covered by two seeds $s_1$ and $s_2$ sequentially. When $br$ is encountered by the second seed $s_2$, the opposite branch of $br$ would be pruned by the LOB strategy because $br$ has already been covered by $s_1$. In contrast, conventional symbolic executors solve the opposite branch of $br$ twice due to different path prefixes for $s_1$ and $s_2$. 

In summary, the LOB strategy can help tailored symbolic executor to focus on uncovered branches, preventing it from being overwhelmed by numerous seeds from the fuzzer.
As shown in Table \ref{tab:existing_architecture_search_strategies} (2nd column), several recently proposed hybrid testing tools, including SymCC, MEUZZ, SYMSAN, and CoFuzz, \textit{etc.}, all use the LOB strategy.

\subsubsection[Observation]{\textbf{Observation}}\label{sec:observation}
We run QSYM, DigFuzz, SymCC, and CoFuzz on prevalent benchmarks for $3\times 24$ hours \footnote{Seeds are from the corpus discussed in Section 5.}. Table \ref{tab:sleep} shows the average time spent by symbolic executors waiting for seeds from the fuzzer.  

\textbf{\underline{Busy Symbolic Executor in QSYM}}. 
Symbolic executor in QSYM exhibits minimal idle time, indicating that it cannot complete all seeds produced by the fuzzer. Thus, the LOB strategy is suitable for QSYM due to its focus on uncovered branches.
DigFuzz \cite{digfuzz_on_qsym} improves QSYM by selecting the hardest path to be fuzzed for symbolic execution, aiming to utilize the ability of symbolic execution to the best in the limited time. In this regard, it is reasonable to explore valuable branches first.

\underline{\textbf{Sleeping Symbolic Executor in SymCC}}.
Although the LOB strategy works well in QSYM, it exhibits drawbacks with more efficient symbolic executors in later proposed hybrid testing systems. For example, SymCC implements symbolic execution in the compilation phase by instrumenting symbolic execution codes at the LLVM IR level. Compared with QSYM, which needs dynamic instrumentation in each run, SymCC needs only once instrumentation and can benefit from compilation optimizations. Thus, SymCC is more efficient than QSYM. Importantly, SymCC also uses the same LOB pruning strategy as QSYM.
As shown in Table \ref{tab:sleep}, SymCC spends 56.42\% of its total time waiting for seeds from the fuzzer.

Note that this sleeping phenomenon is not unique to SymCC. The subsequently proposed hybrid testing system, CoFuzz, which introduces a new seed scheduling method and incorporates sampling into symbolic execution, also spends 37.61\% of its total time waiting for seeds. This suggests that seed scheduling or integrating sampling cannot fully utilize the capabilities of symbolic execution. Other hybrid testing tools (\textit{e.g.}, SYMSAN), which improve symbolic computation efficiency but still use LOB strategy, have the same issue. 

\begin{table}[t] 
        \centering 
\caption{ Average sleeping time (in hours) of symbolic execution during $3\times24$-hour runs of hybrid fuzzers.\textquoteleft N\textquoteright\ means unsupported.}
\label{tab:sleep}     
        
        \begin{tabular}{lccccc}
   
            \toprule
            Program & QSYM & DigFuzz & CoFuzz& SymCC   &  \method\ (our)  \\
            \noalign{\smallskip}
            \hline
            \noalign{\smallskip}
             objdump   & 0.00& 0.00 & 0.00 & \underline{0.00} & \underline{0.00}\\
             libjpeg      & 0.00 & 0.00 &0.00 & \underline{0.00}  & \underline{0.00} \\
             tcpdump      & 0.00& 0.00 & 11.15 &9.46    & \underline{0.00} \\
             libarchive   & 0.00& 0.00 & 5.57& 21.83  & 0.12 \\
             pngimage     &0.04 & 10.16 & 18.53 & 20.38   & 10.38 \\
             jhead     & 1.92 & 2.32 &22.93 &23.10   &7.10\\
             pngfix     & 0.02 & 1.92 & 22.30 & 23.43   & 13.56 \\ 
             libxml2        &0.00& 0.00 & 0.70 & 16.18   & \underline{0.00} \\
             openjpeg   & 0.00& 0.00& 0.00 & \underline{0.00}  & \underline{0.00} \\  
             readelf      & 0.00&0.00 & 0.00 & 12.72  & \underline{0.00}  \\
             gdk      & 0.00& 0.00 &  21.38 & 22.44  & 12.28 \\
             nm       & 0.00& 0.00 & 8.68 & 23.55   & 1.91 \\
             strip          & 0.00  & 0.00 & 4.38 &\underline{0.00} &\underline{0.00}\\
            imginfo      &0.03 &1.31 & 10.76 & 7.43  & 0.44 \\
            cyclondds      & 0.00& 0.00 & N & 22.60   & 0.56 \\
                         \noalign{\smallskip}\hline
            \textbf{average} & 
                \makecell{0.13 \\ 0.54\%} & 
                \makecell{1.05 \\ 4.36\%} & 
                \makecell{9.03 \\ 37.61\%} & 
                \makecell{13.54 \\ 56.42\%} & 
                \makecell{3.09 \\ 12.88\%} \\
            \bottomrule
        \end{tabular}
\end{table}

\subsubsection[Analysis and Our Solution]{\textbf{Analysis}}

The prevalent LOB strategy over-prunes branches and leads to the sleeping issue of symbolic executors.
However, this issue cannot be resolved under the architecture of the current tailored symbolic executors. 
The underlying reasons are twofold.
\begin{enumerate}
	\item First, tailored symbolic executors discard the execution tree and maintain only branch coverage for efficiency. Consequently, it cannot distinguish different path contexts for the same branch statement. Therefore, its design principle substantially restricts the search and pruning strategies used in tailored symbolic executors. Seed scheduling methods \cite{digfuzz_on_qsym, MEUZZ} aim to explore valuable seeds first and are orthogonal to pruning strategies. So these studies cannot mitigate the sleeping issue in symbolic execution either.
	 
	\item Second, tailored symbolic executors typically employ a bitmap to store branch coverage efficiently. However, the bitmap may have severe hash collisions (reaching up to 75\%) in practice \cite{Coll_AFL} because multiple branches may be mapped to the same index. This phenomenon causes additional branches to be over-pruned.
\end{enumerate}

\subsubsection{\textbf{Our Solution}}

A natural question arises: \textit{can we leverage the sleeping time in symbolic execution to explore the branches pruned by the LOB strategy, while maintaining the high efficiency of tailored symbolic executors}? 
In this paper, we argue that the sleeping issue in tailored symbolic executors cannot be addressed solely by a single technique, such as seed scheduling or redesigning hash functions. The architecture of tailored symbolic executors confines several critical components, including the coordination mechanism, search and pruning strategies, \textit{etc}.
We believe that addressing the sleeping issue of symbolic execution requires a redesign of the overall architecture.

To this end, we propose a new symbolic executor for hybrid testing that lies between conventional and tailored symbolic executors.
As illustrated in the middle of Fig. \ref{fig:spectrum}, our tool incorporates a sophisticated coordinator that bridges the fuzzer and a tailored symbolic executor. The coordinator maintains a lightweight execution tree to track path coverage and other essential information, while omitting symbolic states used in conventional symbolic executors. 
Based on the execution tree, we can distinguish different path prefixes of the same branch and keep all paths on the execution tree.
This can avoid over-pruning branches in the LOB strategy.  
We can also design flexible search strategies similar to those in conventional symbolic executors. 
Meanwhile, we still use a tailored symbolic executor for single-path symbolic execution for efficiency.
Overall, our method achieves a balance between flexible conventional symbolic executors and scalable tailored symbolic executors, integrating the advantages of both approaches.

\subsection{Motivation 2: Where to Sample?}

We notice that two existing hybrid testing tools that integrate sampling, PANGOLIN and CoFuzz, replace constraint solving with sampling entirely during symbolic execution. These tools do not consider the test redundancy issue of sampling. 
Here, we use an example to demonstrate why sampling should be applied at appropriate branches.
Fig. \ref{fig:src} shows code snippets from the libxml2 library, which contain the following three branches.

\begin{enumerate}[(1)]
    \item \textbf{Branch $b_1$ }(line 1). The condition of $b_1$ is $\phi_1=\langle \mathsf{val >= 0x80}\rangle$, where $\mathsf{val}$ is a 32-bit integer. Suppose that the inputs generated in fuzzing follow a uniform distribution  $U\sim[-2^{31},2^{31}-1]$. The probability that $b_1$ is covered by a single random mutation on $\mathsf{val}$ is approximately 0.5. In our experiments, AFL can cover $b_1$  within 2 seconds with the initial seed ``0000''. Therefore, there is no need to invoke costly symbolic execution to explore $b_1$.
    
    \item \textbf{Branch $b_2$ }(line 5). The condition of $b_2$ is $\phi_2 = \langle \mathsf{\texttt{CUR\_PTR[0]}=\texttt{`<'} \cdots \wedge \texttt{CUR\_PTR[8]}=\texttt{`E'}  \wedge l = 1}$$\rangle$. The probability of satisfying $\phi_2$ through random mutation is $2^{-104}$. AFL cannot cover $b_2$ in 3 hours using zero seed, whereas $\phi_2$ can be solved almost instantly by a modern SMT solver such as Z3 (within 2 microseconds).
    Since there is only one path after $b_2$, solving once at $b_2$ is sufficient to generate an input that covers line 7. 
    Thus, it is unnecessary to perform expensive sampling at branches where additional inputs cannot bring new coverage or trigger new crashes.
    However, both existing tools integrating sampling, CoFuzz and PANGOLIN, would still apply sampling at $b_2$ once $b_2$ is identified as a hard branch for fuzzing.

    \item  \textbf{Branch $b_3$ }(line 10). The condition of $b_3$ is $\phi_3 = \langle\mathit{domain}=23 \wedge \mathit{in}[0]=\mathtt{0xEF} \wedge \mathit{in}[1]=\mathtt{0xBB} \wedge \mathit{in}[2]=\mathtt{0xBF} \wedge (\mathit{c} \& \mathtt{0xFC00}) \neq \mathtt{0xD800}\rangle$. The probability of satisfying $\phi_3$ through random mutation is $2^{-56}$. So $b_3$ should also be handled by symbolic execution.
    Importantly, branch $b_3$ dominates a large portion of the program (lines 11$\sim$16, many codes are omitted). Different values of $\mathsf{c}$ can lead to different paths below $b_3$. So sampling at $b_3$ can generate multiple inputs that cover different paths along $b_3$. Consequently, $b_3$ is suitable for sampling. 

\end{enumerate}

\textbf{Our Solution}. 
We believe that sampling should be applied at hard-to-cover branches with high expected benefit. 
In this paper, we propose a set of principles for integrating fuzzing, symbolic execution, and sampling. Following these principles, we design a coordination mechanism incorporating different techniques on top of the aforementioned architecture. Experimental results demonstrate the effectiveness of our method.

\begin{figure}[t!]
  \setlength{\intextsep}{0pt} 
  \setlength{\textfloatsep}{0pt} 
  \setlength{\floatsep}{0pt}    
  \setlength{\abovecaptionskip}{-4pt} 
  \setlength{\belowcaptionskip}{0pt} 
  \centering
\begin{lstlisting}[basicstyle=\fontsize{8pt}{8pt}\ttfamily, lineskip={0pt}]
if (val >= 0x80) { /*b1*/
  return xmlCopyCharMultiByte (val);
}

if(CMP9(CUR_PTR,(*@'<'@*),(*@'!'@*),(*@'D'@*),(*@'O'@*),(*@'C'@*),(*@'T'@*),(*@'Y'@*),(*@'P'@*),(*@'E'@*)) && (l == 1))/*b2*/
{//CMP9: verifies 9-char match at CUR_PTR
  b[i++] = v;
}

if(domain==XML_FROM_VALID && in[0]==0xEF && in[1]== 0xBB && in[2]==0xBF && (c&0xFC00)!=0xD800)/*b3*/ 
{//XML_FROM_VALID=23
  if (c < 0x80)         { *out++=  c; bits= -6; }
  else if (c < 0x800)   { *out++= ((c >>  6) & 0x1F) | 0xC0; bits= 0;}
  else if (c < 0x10000) { *out++= ((c >> 12) & 0x0F) | 0xE0; bits= 6;}
  else                  { *out++= ((c >> 18) & 0x07) | 0xF0; bits=12;}
  /* many other uncovered codes ... */
}
\end{lstlisting}
\vspace{-0.1ex}
\captionof{figure}{A motivating example in libxml2}
\label{fig:src}
\end{figure}

\section{Method} \label{sec:method}
 \begin{figure*} [t] 
	\centering
	\includegraphics[width=0.8\linewidth]{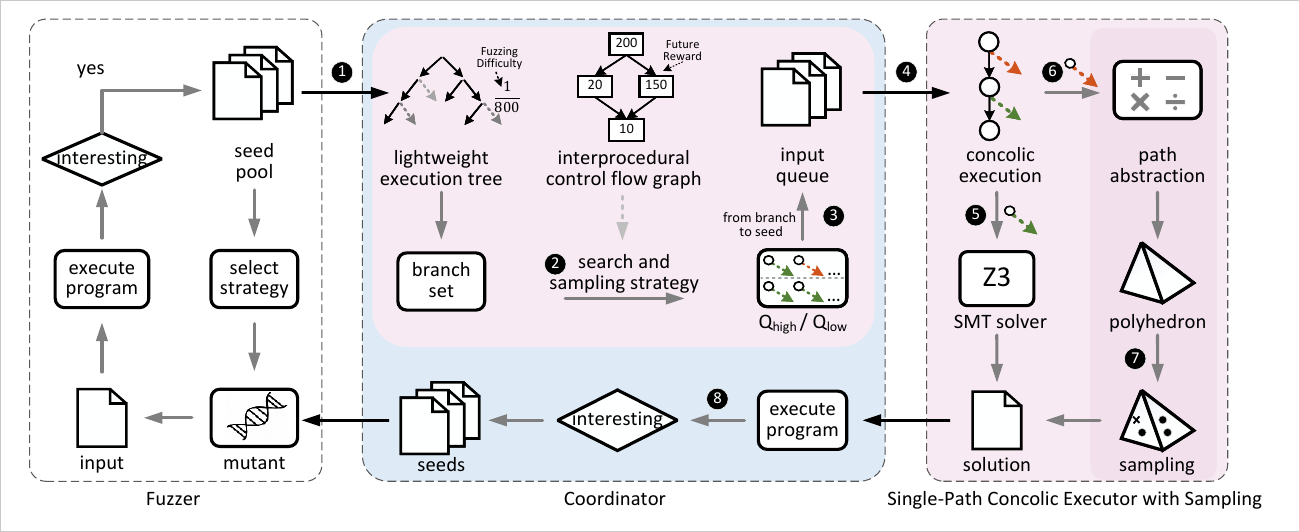}
	\caption{Architecture of our hybrid testing platform.}
	\label{fig:arch}
\end{figure*}
In this section, we first illustrate the architecture of hybrid testing. Next, we present the coordination mechanism for integrating different techniques. Finally, we discuss how to integrate sampling in symbolic execution.
\subsection{Overall Architecture} \label{sec:overview}


We propose a novel hybrid testing architecture. As illustrated in Fig. \ref{fig:arch}, the hybrid testing architecture consists of three components: a fuzzer (left), a symbolic executor (right), and a coordinator in the middle. Our work focuses on the coordinator and symbolic executor (colored regions in Fig. \ref{fig:arch}). Existing fuzzing works can be seamlessly integrated into this framework.

\textbf{Coordinator}. 
Unlike prior tools that simply exchange seeds between fuzzing and symbolic execution \cite{QSYM, SymCC}, our coordinator is more sophisticated. 
First, the coordinator maintains a lightweight execution tree and the interprocedural control flow graph (ICFG). In collaboration with a tailored symbolic executor (right side of Fig. \ref{fig:arch}), it realizes a full-fledged yet efficient symbolic executor.
Second, the coordinator supports flexible search strategies, instantiates our proposed principles for integrating fuzzing, symbolic execution, and sampling, addressing the ``where to sample'' problem.

\textbf{Lightweight Execution Tree}. In conventional symbolic executors, the execution tree typically contains various information, including path conditions and symbolic states, to enable flexible state forking and search strategies. In contrast, each node in our lightweight execution tree stores only pointers to its successors and one pointer to the corresponding branch in the ICFG.
Such a concise design can improve the scalability of the system while preserving the flexibility of conventional symbolic executors.

In our system, the lightweight execution tree serves the following roles.

\begin{enumerate}[(1)]
    \item \textit{Avoiding over-pruning branches}. The execution tree can distinguish different path prefixes for the same branch statement. 
    This retains more opportunities to explore different program states and discover more crashes.
    Therefore, we can avoid over-pruning branches in LOB strategy.
    \item \textit{Supporting advanced search strategies}. Conventional symbolic executors maintain an execution tree to track path coverage, which supports precise path analysis and flexible search strategies (\textit{i.e.}, prioritizing open branches on the tree according to specific testing goals). 
    In contrast, tailored symbolic executors typically depend on seed scheduling and cannot employ fine-grained search strategies to prioritize branches without contexts.
    \item \textit{Improving scalability}. The execution tree is lightweight and can be extended to considerable depth in practice. For instance, in our experiments, we set a depth bound of 15,000, which is difficult to achieve in conventional symbolic executors. Thus, we can track more paths.
\end{enumerate}

As illustrated in Fig. \ref{fig:spectrum}, our system resides in the middle of the spectrum. 
It combines the precise path analysis capability of conventional symbolic executors and the scalability of tailored symbolic executors.

\textbf{Tailored Symbolic Executor With Sampling}. 
The right side of Fig. \ref{fig:arch} illustrates the tailored symbolic executor enhanced with sampling. Similar to several existing hybrid testing tools \cite{SymCC, QSYM}, the tailored symbolic executor in our system also operates as a single-path executor invoked iteratively by the coordinator for each seed. Our system differs from existing tools in two aspects. First, the coordinator, rather than the symbolic executor, decides which branch to solve during execution. Second, we integrate sampling into symbolic execution at selected appropriate branches.

\textbf{Workflow}.
The workflow of the hybrid testing system consists of the following key steps.
\ding{182} \textit{Constructing/updating lightweight execution tree}. The coordinator collects seeds from the fuzzer and constructs/updates a lightweight execution tree. 
\ding{183} \textit{Search and sample strategy}. Open branches on the execution tree are prioritized into two priority queues based on two criteria: (i) the difficulty of being covered by the fuzzer, and (ii) the potential reward estimated using a pre-built ICFG. The coordinator then determines which open branches should be solved ({\color{ForestGreen} green}) or sampled ({\color{BrickRed}red}) following several principles (discussed later).
\ding{184} \textit{From branch to seed priority}. According to branches in priority queues, the highest-priority input is selected for symbolic execution.
\ding{185} \textit{Symbolic execution}. The coordinator sends the selected seed to the tailored symbolic executor for concolic execution. 
\ding{186} \textit{Constraint solving}. If the branch is marked for constraint solving ({\color{ForestGreen} green}), its path condition is sent to a constraint solver.

Otherwise, the path condition is abstracted into a polyhedron at step \ding{187} \textit{Path abstraction},  and then sampled to generate a set of inputs that are likely to satisfy the original condition at step  \ding{188} \textit{Sampling}.
\ding{189} \textit{Synchronizing seeds back to fuzzer}. Input generated through solving or sampling is synchronized back to the fuzzer if it contributes new coverage.

\subsection{Coordination Mechanism}
Building on the proposed architecture, we design a coordination mechanism to address the issues discussed in Section \ref{sec:problem}.
This subsection first introduces the design principles for integrating different techniques, then presents the coordination algorithm that instantiates these principles.

\subsubsection[Our Solution: Principles of Hybrid Testing]{\textbf{Principles of Hybrid Testing}}

\begin{figure} [!t] 
	\centering
	\includegraphics[width=0.9\linewidth]{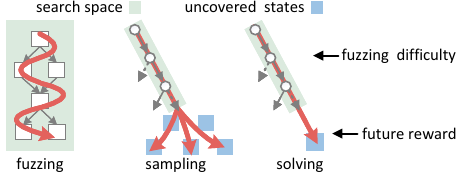}
	\caption{Comparison of three techniques.}
	\label{fig:solutionspace}
\end{figure}

Fig. \ref{fig:solutionspace} illustrates the differences between fuzzing, symbolic execution, and sampling. Fuzzing generates new inputs through low-cost random mutations, but lacks control over the search direction and often incurs high test redundancy. Symbolic execution can precisely explore target branches but depends on expensive symbolic computation and constraint solving. Sampling generates multiple solutions along a specific branch. Although once sampling is more expensive than a single constraint solving query, the amortized cost per input produced by sampling is between fuzzing and constraint solving. 
Similarly, sampling also has the test redundancy issue, because inputs sampled from a branch may not necessarily yield new coverage.

\begin{figure}[t] 
	\centering
	\includegraphics[width=0.9\linewidth]{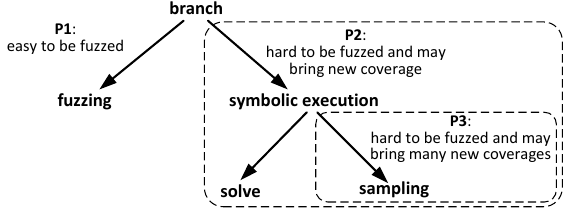}
	\caption{Principles of hybrid testing.}
	\label{fig:principles}
\end{figure}

Suppose three techniques are performed on a target program.
In this paper, we define the cost of a testing method as the CPU time of running this method and the effectiveness of a testing method as the attained testing coverage or discovered crashes.
We note the costs of fuzzing, constraint solving, and sampling as $C_{\text{\textit{fuzz}}}$, $C_\text{\textit{solve}}$, and $C_\text{\textit{sample}}$, respectively. The effectiveness of fuzzing, constraint solving, and sampling is represented by $E_\text{\textit{fuzz}}$, $E_\text{\textit{solve}}$, and $E_\text{\textit{sample}}$, respectively. 
We claim that the optimal hybrid testing architecture should minimize the following cost-effectiveness ratio (CER).
$$
CER=\frac{C_\textit{\textit{fuzz}}+C_\textit{\textit{solve}}+C_\textit{sample}}{E_\textit{fuzz}+E_\textit{solve}+E_\textit{sample}}
$$

In practice, the costs of different techniques are related to the program characteristics and the underlying testing methods. The effectiveness is also determined by multiple factors, including the testing goal (coverage or crash discovery), the program structure, \textit{etc}. Thus, both the cost and effectiveness are hard to measure due to the natural complexity of each technique. 
Nevertheless, we can still step towards the optimal architecture with minimum CER.
As shown in Fig. \ref{fig:principles}, considering the effectiveness and costs of the three techniques, we propose the following three principles for hybrid testing. 
\begin{itemize}
	\item \bm{$P_1$}: Branches that are easy to be covered by (semi-) randomly generated inputs should be handled by fuzzing, because covering these branches with fuzzing incurs minimum overhead ($C_\textit{\textit{fuzz}}$). It is unnecessary to perform constraint solving or sampling on these branches (higher $C_\textit{\textit{solve}}$ or $C_\textit{\textit{sample}}$).
	
	\item \bm{$P_2$}: Branches that are difficult for fuzzing and also may lead to new coverage should be explored by symbolic execution.  Modern constraint solvers equipped in symbolic executors can solve highly complex constraints that are hard to satisfy by random mutation\footnote{Symbolic executors often use a timeout threshold to control the time budget on problems (\textit{e.g.}, complex nonlinear constraints) that exceed the capability of modern constraint solvers.}. Once these branches are solved by symbolic execution, the generated input can be mutated by fuzzing further to achieve more coverage and crashes. Therefore, this can reduce the cost of fuzzing ($C_\textit{fuzz}$) and increase the effectiveness of symbolic execution ($E_\textit{solve}$) and fuzzing ($E_\textit{fuzz}$).
	
	\item \bm{$P_3$}: (a subcase of \bm{$P_2$}): For branches that are difficult for fuzzing to cover and are likely to expose numerous new paths, sampling can be applied. This can gain a high effectiveness $E_\textit{sample}$. Since sampling requires symbolic execution to obtain the path condition beforehand, we consider sampling (\bm{$P_3$}) as a subcase of symbolic execution (\bm{$P_2$}).  
\end{itemize}

These principles are influenced by multiple factors in practice. For example, in the SOTA hybrid testing systems that run fuzzer and symbolic executor in parallel, principle $P_1$ is naturally realized by the fuzzer. The coordinator analyzes branches that have not been explored by fuzzing and invokes the symbolic executor to implement $P_2$ and $P_3$.

\subsubsection[coordination mechanism]{\textbf{Coordination Algorithm}}

Algorithm \ref{alg:coordinator} shows the details of the coordination algorithm. 
The whole procedure is a while loop until the time resource is exhausted. Overall, the loop body consists of two halves. The first half runs periodically and conditionally. It maintains a lightweight execution tree and implements a search and sample strategy. The second half selects seeds from branch queues and runs them via sampling-enhanced concolic execution.

\textbf{The First Half: Maintaining Execution Tree and Integrating Search and Sampling Strategies}. In the beginning, the algorithm initializes two-tier priority queues, $Q_{high}$ and $Q_{low}$, for sorting open branches. Variables $last$ and $current$ are used to store times.
The first half of the loop body (lines \ref{alg:sync} $\sim$ \ref{alg:sortqueue}) runs when a specified period $\mathcal{T}$ is reached or the queue $Q_{high}$ is empty. At first,  function $\mathsf{syncSeed}$ (line \ref{alg:sync}) synchronizes seeds from the fuzzer. Function $\mathsf{updateExeTree}$ re-executes these seeds on a separate instrumented binary and collects their execution paths to construct or update the lightweight execution tree. Meanwhile, all the open branches on the tree are identified and then saved into a set $\textit{$B_{\text{ob}}$}$  (line \ref{alg:exetree}). 
Line \ref{alg:sync} and \ref{alg:exetree} together implement step \ding{182} (\textit{Constructing lightweight execution tree}) in Fig. \ref{fig:arch}.

{\scriptsize
\begin{algorithm}[!t]
\footnotesize
  \caption{Coordination Algorithm of \method}  \label{alg:coordinator}
  \KwIn{program $\mathcal{P}$, parameter $\lambda$, thresholds ${\delta}$, ${\gamma}$}
  $Q_{high} = \emptyset$, $Q_{low} = \emptyset$\hspace{\stretch{1}}\textcolor{darkgray}{\scriptsize/*Initialize two empty queues*/}\\
  \textit{last}, \textit{current} = $\mathsf{getTime}$()\\
  \While{true}{
  	\textit{current} = $\mathsf{getTime}$()\\
    \textcolor{darkgray}{\scriptsize/*{\color{ForestGreen}First half}: synchronize seeds and maintain lightweight execution tree, {\color{ForestGreen} run periodically}*/}\\
    \If{\textit{current}$-$\textit{last} $\geq$ $\mathcal{T}$ \textbf{or} $Q_{high} = \emptyset$}{
      $\mathsf{syncSeed}()$\label{alg:sync}\hspace{\stretch{1}}\textcolor{darkgray}{\scriptsize/*step \ding{182}, synchronize seeds from fuzzer*/}\\
      $B_{\text{ob}} = \mathsf{updateExeTree}()$\label{alg:exetree}\hspace{\stretch{1}}\textcolor{darkgray}{\scriptsize/*step \ding{182}, update execution tree*/}\\
      \textcolor{darkgray}{\scriptsize/*Step \ding{183}, search and sample strategy, see Section \ref{sec:strategySE}*/}\\
      $(Q_{high}, Q_{low}) = \mathsf{searchAndSampleStrategy}(B_{\text{ob}}, \lambda, \mathcal{\delta}, \mathcal{\gamma})$ \label{alg:sortqueue} \\
      
      \textit{last}=$\mathsf{getTime}$()\\
    }
    \textcolor{darkgray}{\scriptsize/*{\color{BrickRed} Second half}: select seed for concolic execution*/}\\
    \eIf{$Q_{high} \neq \emptyset$ \textbf{or} $Q_{low} \neq \emptyset$}{\label{alg:checkempty}
      \eIf{$Q_{high} \neq \emptyset$}{
      	\textcolor{darkgray}{\scriptsize/*Step \ding{184}, from branch to seed, see Section \ref{sec:branchtoseed}*/}\\
      	$(seed, {action}_{seed}) = \mathsf{mostPrioritizedSeed}(Q_{high})$\label{alg:getnexthigh}
      }{
      	$(seed, {action}_{seed}) = \mathsf{mostPrioritizedSeed}(Q_{low})$\label{alg:getnextlow} 
      }
      \textcolor{darkgray}{\scriptsize/*Step \ding{185}$\sim$\ding{188}, symbolic execution, see Section \ref{sec:CE_Sampling}*/}\\
      $\textit{res} = \mathsf{concolicWithSampling}(\mathcal{P}, seed, {action}_{seed})$\label{alg:concolic}\\
      $\mathsf{syncBackSeed}(\textit{res})$\\ \label{alg:syncback} 
    }{
      $\textit{wait}$\hspace{\stretch{1}}\textcolor{darkgray}{\scriptsize/*wait for new seed*/}\label{alg:sleep}          
    }\label{alg:over}
    \If{timeout}{break}
  }
\end{algorithm}
}

Function $\mathsf{searchAndSampleStrategy}$ instantiates the hybrid testing principles discussed earlier.
Each open branch is queued into the high-priority queue $Q_{high}$ or the low-priority queue $Q_{low}$ according to its priority.

Each element in $Q_{high}$ and $Q_{low}$ is a triple $\langle ob, S, act\rangle$, where $ob$ is an open branch, $S$ is the priority of $ob$, $act$ is a flag with value $\mathsf{SOLVE}$ or $\mathsf{SAMPLE}$, indicating whether \textit{constraint solving} or \textit{sampling} at $ob$ during symbolic execution.
Overall, function $\mathsf{searchAndSampleStrategy}$ implements step \ding{183} (\textit{search and sample strategy}) in Fig. \ref{fig:arch}. We will elaborate on the function $\mathsf{searchAndSampleStrategy}$ and the priority queues in Subsection \ref{sec:strategySE} in detail.

In practice, it may take tens of seconds to sort all branches for large programs. For example, the execution tree contains more than 8,000 seeds and 10 million open branches when testing \textit{tcpdump}. It is inappropriate to sort branches each time when one seed is executed by symbolic execution.
To avoid frequent branch sorting, we configure the first half to execute periodically.
In our experiments, we set the period $\mathcal{T}$ as 30 minutes. We observe that the first half is executed frequently during the early phase of testing, when $Q_{high}$ is not large. As testing progresses, more seeds are added to the execution tree, and the size of $Q_{high}$ increases. In this case, the first half is invoked after the period $\mathcal{T}$ elapses.


\textbf{The Second Half: Concolic Execution With Sampling}. The second half of the while loop (line \ref{alg:checkempty} $\sim$ \ref{alg:syncback}) explores open branches using symbolic execution and generates seeds. If the high priority queue $Q_{high}$ is not empty, function $\mathsf{mostPrioritizedSeed}$ (line \ref{alg:getnexthigh}) implements step \ding{184} (\textit{From branch to seed priority}).
It selects the most prioritized branch $\textit{$ob$}$ from $\textit{$Q_{high}$}$, then selects $ob$'s corresponding $seed$ for further symbolic execution.  In Section \ref{sec:branchtoseed}, we will elaborate on the correspondence between open branches and seeds. Note that a seed may associate multiple open branches. Function  $\mathsf{mostPrioritizedSeed}$ analyzes all the open branches associated with $seed$, but skips open branches residing in $Q_{low}$. Such a designation can force the system to explore open branches in $Q_{high}$ first.
Only when $Q_{high}$ is empty, open branches in $Q_{low}$ would be explored (line \ref{alg:getnextlow}).

Besides the selected $seed$, function  $\mathsf{mostPrioritizedSeed}$ also returns  $\text{\textit{action}}_{seed}$, a sequence of actions corresponding to the sequence of open branches $\{ob_1, \cdots, ob_n\}$ associated with $seed$. 
For example, $action_{seed}=\langle\mathsf{SOLVE},\mathsf{SOLVE},\mathsf{SAMPLE}\rangle$
indicates that when executing $seed$ (line \ref{alg:concolic}), the symbolic executor solves at the first two branches, and samples at the third branch. 
In this way, the coordinator implements the hybrid testing principles discussed earlier.
After executing the seed, all the generated test cases are synchronized back to the fuzzer if they contribute to coverage or trigger crashes (line \ref{alg:syncback}). 
In Section \ref{sec:CE_Sampling}, we will elaborate on the details of symbolic execution enhanced with sampling.

Finally, the second half would sleep for a short period if both $Q_{high}$ and $Q_{low}$ are empty (line \ref{alg:sleep}).

Note that the first half synchronizes seeds from the fuzzer as soon as $Q_{high}$ is empty, even when branches in $Q_{low}$ are not processed yet. This guarantees that branches with low benefits are always explored with a low priority.

Finally, if both $Q_{high}$ and $Q_{low}$ are empty, the coordinator waits for the fuzzer to produce the next new seed (line \ref{alg:sleep}).

\subsection{Search and Sampling Strategy}\label{sec:strategySE}
We can treat the three techniques in our system as different search methods over the path space. 
To implement the hybrid testing principles discussed earlier, the search strategy should decide: (1) which branches are explored by symbolic execution, (2) the priority of branches and seeds, and (3) which branches should be sampled.  
Therefore, the system must determine which branches are difficult for fuzzing to cover and whether they are likely to expose a large number of unexplored paths.
In Algorithm~\ref{alg:coordinator}, these principles are implemented in the function $\mathsf{searchAndSampleStrategy}$.

\begin{algorithm}[!t]
\footnotesize
\setlength{\abovecaptionskip}{2pt} 
\SetAlgoLined
\caption{$\mathsf{searchAndSampleStrategy}$($B, \lambda, \mathcal{\delta}, \mathcal{\gamma}$): search strategy of symbolic execution}
\label{alg:searchAndSampleStrategy}
\KwIn{set $B$ of $ob$, parameter $\lambda$, thresholds $\mathcal{\delta}, \mathcal{\gamma}$}
\KwOut{two prioritized branch queues $(Q_{high}, Q_{low})$}
\SetKw{KwSort}{sort}

\ForEach{$ob_i \in B$}{
	\textcolor{darkgray}{{\scriptsize/*estimated branch probability as difficulty*/}}\\
    $D_i = \mathsf{fuzzingDifficulty}(ob_i)$\\ \label{alg:difficulty}
    $R_i = \mathsf{futureReward}(ob_i)$ \label{alg:reward}
    
    \If{$D_i < \mathcal{\delta}  \ and \ R_i>\mathcal{\gamma}$}{ \label{alg:solveorsample_start}
    	$action_i$  = $\mathsf{SAMPLE}$  \textcolor{darkgray}{{\color{BrickRed}\scriptsize/*sample at hard and valuable branches*/}}\label{alg:samplehere}\\
    }
    \Else{
    	$action_i$  = $\mathsf{SOLVE}$ \\ 
    }\label{alg:solveorsample_end}
    $S_i = \lambda \times normalize(D_i) + (1-\lambda)\times normalize(R_i)$ \label{alg:score}
    
    \If{$R_i > 0$}{
        $Q_{high}.\text{insert}(\langle ob_i, S_i, action_i\rangle)$ \label{alg:inserthigh}
    }
    \Else{
        $Q_{low}.\text{insert}(\langle ob_i, S_i, action_i\rangle)$
    }
}

\KwRet{$(Q_{high}, Q_{low})$}
\end{algorithm}

Algorithm \ref{alg:searchAndSampleStrategy} describes the procedure of the function $\mathsf{searchAndSampleStrategy}$.
The input consists of a set $B$ of open branches, a parameter $\lambda$, and thresholds $\delta$, $\gamma$. For each open branch $ob_i \in B$, 
function $\mathsf{fuzzingDifficulty}$ estimates the probability of  $D_i$ as the fuzzing difficulty at line \ref{alg:difficulty} (A lower $D_i$ implies a higher difficulty). Function $\mathsf{futureReward}$ approximates the number of uncovered statements potentially reachable from $\textit{$ob_i$}$ (line \ref{alg:reward}). Based on these two factors, lines \ref{alg:solveorsample_start} $\sim$ \ref{alg:solveorsample_end} determine whether to perform constraint solving or sampling at $ob_i$. If the fuzzing difficulty $D_i < \mathcal{\delta}$ and the future reward $R_i>\mathcal{\gamma}$, the algorithm performs sampling at $ob_i$. Otherwise, it performs constraint solving. This strategy ensures that sampling is applied only to branches that are difficult for fuzzing yet still promise high coverage gains. We discuss the estimation of fuzzing difficulty and future reward later.

For each open branch, the algorithm computes a priority score $S_i$ according to $D_i$ and $R_i$ (line \ref{alg:score}), where the parameter $\lambda$ controls the contributions of these two factors.
For an easy-to-cover open branch $ob_i$ with a low priority in symbolic execution, the fuzzer may cover $ob_i$ before symbolic execution explores it. This naturally enforces symbolic execution to explore hard-to-cover branches for fuzzing (principle $\bm{P_1}$).

Finally, if $R_i>0$, \textit{$ob_i$} is inserted into $Q_{high}$ (line \ref{alg:inserthigh}). Otherwise, \textit{$ob_i$} is inserted into $Q_{low}$. 
Our method does not prune branches in $Q_{low}$ for two reasons. First, these low-priority branches may still contribute to new path coverage. Second, they can prevent symbolic execution from sleeping when all high-priority branches have been explored. Note that such a fine-grained search strategy cannot be realized in tailored symbolic executors because they lack global path coverage.

To prevent unnecessary loop unrolling, we only put the first and the last branches produced by loop unrolling into $Q_{high}$. Remaining occurrences of loop-induced branches are placed into $Q_{low}$, regardless of whether their future rewards are zero.

In the following, we discuss how to compute the fuzzing difficulty $D_i$ and future reward $R_i$.

\subsubsection{Fuzzing Difficulty }\label{sec:fuzzdiff}
An ideal estimation of the fuzzing difficulty of covering a branch needs to compute the size of the corresponding input space. However, this is more difficult than symbolic execution when testing real-world programs. In this work, we adopt the method proposed in Digfuzz \cite{Dig_Fuzz} to estimate the fuzzing difficulty of a branch. 
Digfuzz uses Monte Carlo method to estimate the fuzzing difficulty of each branch.
In the fuzzing process, the larger the input space of a branch, the more inputs traverse the branch. When the execution reaches a conditional statement, the probability of covering each of its two branches is estimated through hit numbers in history.
For example, when a branch $br$ and its opposite branch have both been covered 10 times, the probability of $br$ is estimated as 0.5.
Formally, the \textit{local probability} of covering a branch \textit{$br$} is estimated by
\begin{equation}
	P_{local}\left(br\right)=\left\{\begin{array}{ll}\frac{{cov}\left(br\right)}{{cov}\left(br\right)+{cov}\left(\bar{br}\right)}, & {cov}\left(br\right) > 0 \\\frac{1}{{cov}\left(\bar{br}\right)}, & {cov}\left(br\right)=0\end{array}\right.
\end{equation}
where \textit{$br$} and \textit{$\bar{br}$} are opposite branches of the same conditional statement, $cov(br)$ is the number of inputs traversing branch \textit{$br$} in the control flow graph \footnote{Note that a single branch $br$ in the control flow graph may occur multiple times in the execution tree. Here, the hit number $cov(br)$ is collected in the CFG rather than the execution tree.}. When $cov(br)>0$, the local probability of $br$ equals the ratio of executions along $br$ and that along both $br$ and $\bar{br}$. If $br$ has not been covered yet ($cov(br) = 0$), the method assigns a small value to $P(br)$ by the \textit{rule of three} in statistics. 

The probability of an open branch $ob_i$ on the execution tree can be estimated as the multiplication of the local probability of each branch $br_j$ on $ob_i$'s path prefix $\langle br_1,\cdots, br_n \rangle$ as follows\footnote{The neighboring items in Equation \ref{eq:path_prob} does not reduce because $P_{local}$ is calculated according to the covered times of branch in CFG.  More details can be referred to \cite{Dig_Fuzz}.}.
\begin{equation}\label{eq:path_prob}
	D_i=P\left(ob_i\right)=\prod_{1\leq j\leq n} P_{local}\left(br_j\right)
\end{equation}

Finally, the normalized fuzzing difficulty of each open branch is the normalized probability
\begin{equation}\label{eq:D}
normalize(D_{i})=\frac{\log P(ob)}{\log P_{min}}
\end{equation}
where $P_{min}$ is the minimal path probability of all open branches.
$normalize(D_i)$ is always not greater than 1.

\subsubsection{Future Reward}\label{sec:reward}
The future reward of \textit{$ob_i$} reflects the value of exploring program parts after \textit{$ob_i$}. This work aims to maximize code coverage within a limited time budget.
So the normalized future reward $R_{i}$ of \textit{$ob_i$} is designed as
\begin{equation}
    normalize(R_{i})=\frac{R_i}{1000} 
\end{equation} 
where $R_{i}$ is the lines of uncovered code that \textit{$ob_i$} can reach. In practice, it is too expensive to calculate $R_{i}$ if considering path feasibility. In this work, we count the lines of uncovered reachable code. 

To implement the above functionality efficiently, we use an adjacency matrix to represent the CFG of each function. Reachable uncovered codes from each block within the same function are tracked in the adjacent matrix during testing. Reachable functions from each block are also stored. Based on these data structures, reachable uncovered codes from each branch can be collected recursively in an efficient way.

\subsection{From Branch to Seed Priority}\label{sec:branchtoseed}
\begin{figure}[t] 
	\centering
	\includegraphics[width=0.20\linewidth]{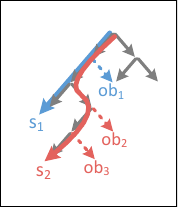}
    \vspace{-2mm}
	\caption{From branch to seed priority.}
    \vspace{-1mm}
	\label{fig:branchTOseed}
\end{figure}
In Algorithm \ref{alg:coordinator}, function $\mathsf{mostPrioritizedSeed}$ (line \ref{alg:getnexthigh} and \ref{alg:getnextlow}) selects the highest-priority open branch $ob$ and returns its corresponding seed for symbolic execution. We adopt a straightforward approach to maintain the correspondence between each open branch $ob$ and its associated seed. Whenever a new seed is synchronized from the fuzzer, all open branches along its path that are newly added to the execution tree are marked as corresponding to that seed.

For example, consider the case shown in Fig. \ref{fig:branchTOseed}. After seed $s_1$ and its path $p_1$ are added to the execution tree, $s_2$ and its path $p_2$ are updated to the tree sequentially. In this process, open branches $ob_2$ and $ob_3$ are associated with $s_2$, while $ob_1$ retains its original association with $s_1$. Consequently, if $ob_2$ has the highest priority, $s_2$ would be selected for symbolic execution in the next step.

Once a seed is selected (line \ref{alg:getnexthigh} of Algorithm \ref{alg:coordinator}), all open branches associated with the seed and currently enqueued in $Q_{high}$ would be explored. In contrast, open branches in $Q_{low}$ will be deferred until $Q_{high}$ is empty. The coordinator records these exploration decisions in the sequence $action_{seed}$ (line \ref{alg:getnexthigh}).
For example, suppose $ob_2$ are enqueued in $Q_{high}$ while $ob_3$ is in $Q_{low}$. When executing seed $s_2$, $ob_2$ is explored while $ob_3$ is skipped.

\subsection{Tailored Concolic Executor With Sampling}\label{sec:CE_Sampling}

\begin{algorithm}[!ht]
\footnotesize
  \caption{$\mathsf{concolicWithSampling}(\mathcal{P}, seed, action_{seed})$}
  \KwIn{$seed$: program $\mathcal{P}$, input $seed$, $action_{seed}$: actions should be taken at open branches along the path of $seed$}
  \KwOut{Solutions $res$}
  \label{alg:pathConstraint}
  $res = \emptyset$\;
  $i = -1$\tcp*{\scriptsize\textcolor{darkgray}{counter for conditional instructions}}
  \ForAll{instructions $ins$ on the path of $seed$}{
      \If{$ins$ is not conditional instruction}{
          update symbolic states\\
      }
      \Else{
          $i = i + 1$\\
          $PC = \mathsf{pathCondition}(ins)$\label{alg:PC}\\
          
          \If{$action_{seed}[i] = \mathsf{SOLVE}$}{
              $res = res \cup SMTSolver(PC)$\\
          }
          \ElseIf{$action_{seed}[i] = \mathsf{SAMPLE}$}{\label{alg:casesample}
              \If{the number of variables in $PC < \mathcal{K}$}{
                  $Poly = \mathsf{PolyhedralAbstraction}(PC)$
                  \label{alg:Polyhedral}\
                  
                  $res = res \cup \mathsf{JohnWalk}(Poly)$\label{alg:Johnwalk}
              }
              \Else{
                  $Rec = \mathsf{IntervalAbstraction}(PC)$\label{alg:interval}\
                  
                  $res = res \cup \mathsf{random}(Rec)$\label{alg:Random}
              }
          }   
      }    
  }
\end{algorithm}

Algorithm \ref{alg:pathConstraint} presents the tailored concolic executor with sampling.
The algorithm takes as input a $seed$ and a sequence $action_{seed}$, which specifies the action to be performed at each open branch along the execution path. For each open branch, the corresponding path condition ($PC$) is collected (line \ref{alg:PC}).
If the action at a branch is $\mathsf{SOLVE}$, an SMT solver is invoked, returning a solution if $PC$ is satisfiable, or an empty set otherwise. If the action is $\mathsf{SAMPLE}$ (line \ref{alg:casesample}), the TaichiPoly method \cite{Taichi} is first used to compute a polyhedral abstraction of the path condition (line \ref{alg:Polyhedral}). Subsequently, John Walk \cite{Jhon_Walk}, a well-established uniform sampling method over polyhedra, is employed to generate a set of samples (line \ref{alg:Johnwalk}).

In our experiments, we observed that computing the polyhedral abstraction becomes costly for complex path conditions. Therefore, when the dimension of a path condition (\textit{i.e.}, the number of variables) exceeds a threshold $\mathcal{K}$, we switch to the interval abstraction method TaichiInt\cite{Taichi}. Although less precise than TaichiPoly, TaichiInt is computationally more efficient and produces a hyperrectangle enclosing the solution space (line \ref{alg:interval}), from which samples can be drawn efficiently. In experiments, we observe the computation time of path abstraction and set the threshold $\mathcal{K}$ to 10.

{\textbf{Sample Size}}. 
The number of samples can influence the testing effectiveness. A large number of samples may introduce redundancy, whereas too few samples may not fully utilize the capabilities of symbolic execution. In practice, determining the optimal sample size is challenging.
To address this, we conduct experiments and observe the relation between sample size and the number of covered paths.
Results show that when the sampling size exceeds 300, the sampled inputs tend to cause testing redundancy. 
Finally, we empirically set the sample size to 300 for polyhedral abstractions and 150 for interval abstractions.

\section{Implementation and Evaluation} \label{sec:ImplEval}

\subsection{Implementation}
We implement our method on top of AFL (v2.56b) and SymCC, and develop a new coordinator in C++ from scratch. For each program under test, we instrument three separate versions: one for fuzzing with AFL, one for symbolic execution, and one for the coordinator to collect the path of each seeds.
For the sampling algorithm, we reimplement the polyhedron abstraction methods TaichiPoly and interval TaichiInt proposed in \cite{Taichi}. We use PPlite \cite{PPlite} and Z3 \cite{Z3} for computing path abstraction.

Overall, the system contains approximately 6,000 lines of C++ code. We name our tool \method\ because it integrates Symbolic execution, Sampling, and Fuzzing.


\textbf{Parameters}.
Algorithm~\ref{alg:searchAndSampleStrategy} relies on $\delta$ and $\gamma$  to control whether to sample and $\lambda$ to  compute branch priority (line~\ref{alg:score}).
We determine their optimal values through grid search.
In the experiments, we set $\lambda = 0.1$ to balance fuzzing difficulty and future reward when computing branch priority, fix $\delta = 10^{-150}$ for all programs, assign $\gamma = 300$ and $80$ for large ($\geq 10$KLOC) and small programs, respectively.

\newcolumntype{C}[1]{>{\centering\arraybackslash}p{#1}}


\subsection{Experimental Setting}
We conduct experiments to evaluate the effectiveness of \method\, aiming to answer the following research questions:

\begin{itemize}
	\item  \textbf{Q1}: Can \method\ outperform SOTA hybrid testing tools in testing efficiency? We use testing coverage and crash numbers to measure testing efficiency. 
    \item  \textbf{Q2}: Can \method\ effectively address the sleeping issue of symbolic execution in hybrid testing?
    \item \textbf{Q3}: How does the sampling strategy affect the testing efficiency?
\end{itemize}
\textbf{Baselines}.
We compare \method\ with two fuzzers, AFL and AFL++, and four hybrid testing tools as follows.

\begin{table}[!t]
\scriptsize
\centering
\caption{{Real-world programs in benchmark.}}
\label{tab:2}       
\begin{tabular}{l p{1.77cm} l p{2.4cm} l}
		\hline
        \noalign{\smallskip}	
		Program & Version &Input & \hspace{0.1cm}Argument &LOC \\
		\noalign{\smallskip}\hline\noalign{\smallskip}
        objdump &binutils-2.28 & ELF & -S @@ &72K\\
        djpeg  &libjpeg-v8a   & JPEG   & @@  &14K	\\
        tcpdump & 5.0.0  & PCAP  & -e -r @@   &47K   \\ 
        bsdtar  &commit-9ebb248 & TAR   & tf @@ &43K	\\
        pngimage   & libpng-1.6.37    & PNG   & @@  &11K  \\ 
        jhead  & jhead-3.08   & JPEG &@@ &2K\\
        pngfix   & libpng-1.6.37    & PNG   & @@  &12K	  \\ 
        xmllint  &libxml2-2.8.0          & XML   & --noout @@        &72K	          \\ 
        opj\_decompress   &openjpeg-2.5.0           &JPEG2  &\tiny{-i @@ -o /tmp/image.pgm} &54K	\\
        readelf &  binutils-2.33.1          & ELF   & \tiny{-agteSdcWw --dyn-syms -D @@}  &29K    \\
        gdk &\tiny{gdk-pixbuf-2.31.1}&JPG&@@ /dev/null &9K\\
        nm  &  binutils-2.33.1          & ELF   & -C -a -l --synthetic @@         &51K        \\ 
        strip    &binutils-2.28  &  ELF  & @@ &64K \\  
          imginfo&jasper-2.0.12&JPG &-f @@  &14K\\
          cyclonedds & commit-53cf7cf4       &IDL&@@ -o /dev/null &9K	\\
		\noalign{\smallskip}

        \hline
	\end{tabular}
\end{table} 
\begin{itemize}
 \item\textbf{AFL} \cite{AFL} is a well-known coverage-guided fuzzing tool.
\item\textbf{AFL++} \cite{afl++} combines various state-of-the-art fuzzing techniques to improve AFL.

\item\textbf{QSYM} \cite{QSYM,qsym_github} is the first to propose a tailored symbolic executor in hybrid testing of binary programs.
 \item\textbf{DigFuzz} \cite{Dig_Fuzz, digfuzz_on_qsym} sends hard branches to a tailored symbolic executor, aiming to make fuzzing and symbolic execution cooperate more complementarily. 
 Since the published code base of DigFuzz \cite{digfuzz_github} is not runnable, 
We reimplement DigFuzz based on QSYM according to its original publication \cite{digfuzz_on_qsym}.
 \item\textbf{SymCC} \cite{SymCC} is one of the most famous hybrid testing tools in recent years \cite{symcc_github}. 
  \item\textbf{CoFuzz} \cite{Co_Fuzz} is the only available hybrid testing tool integrating sampling. We use its original code base \cite{cofuzz_github} to conduct experiments.
\end{itemize}

\begin{figure*}[ht]
	\centering
	\begin{subfigure}{0.195\textwidth}  
		\centering  
		\includegraphics[width=\textwidth]{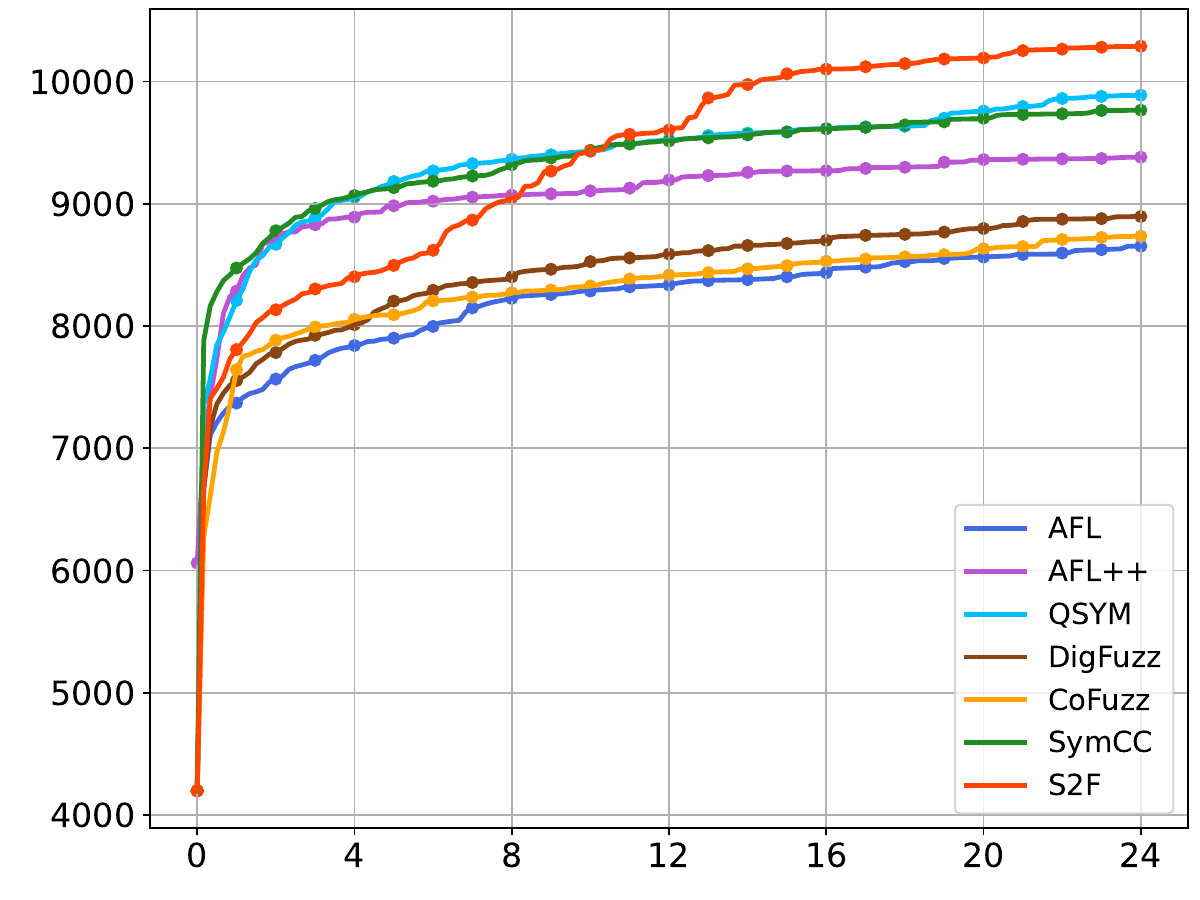}  
		\captionsetup{aboveskip=-1pt, belowskip=-1pt}
		\caption{objdump}  
		\label{fig:objdump}  
	\end{subfigure} 
	\hfill  
	\begin{subfigure}{0.195\textwidth}  
		\centering  
		\includegraphics[width=\textwidth]{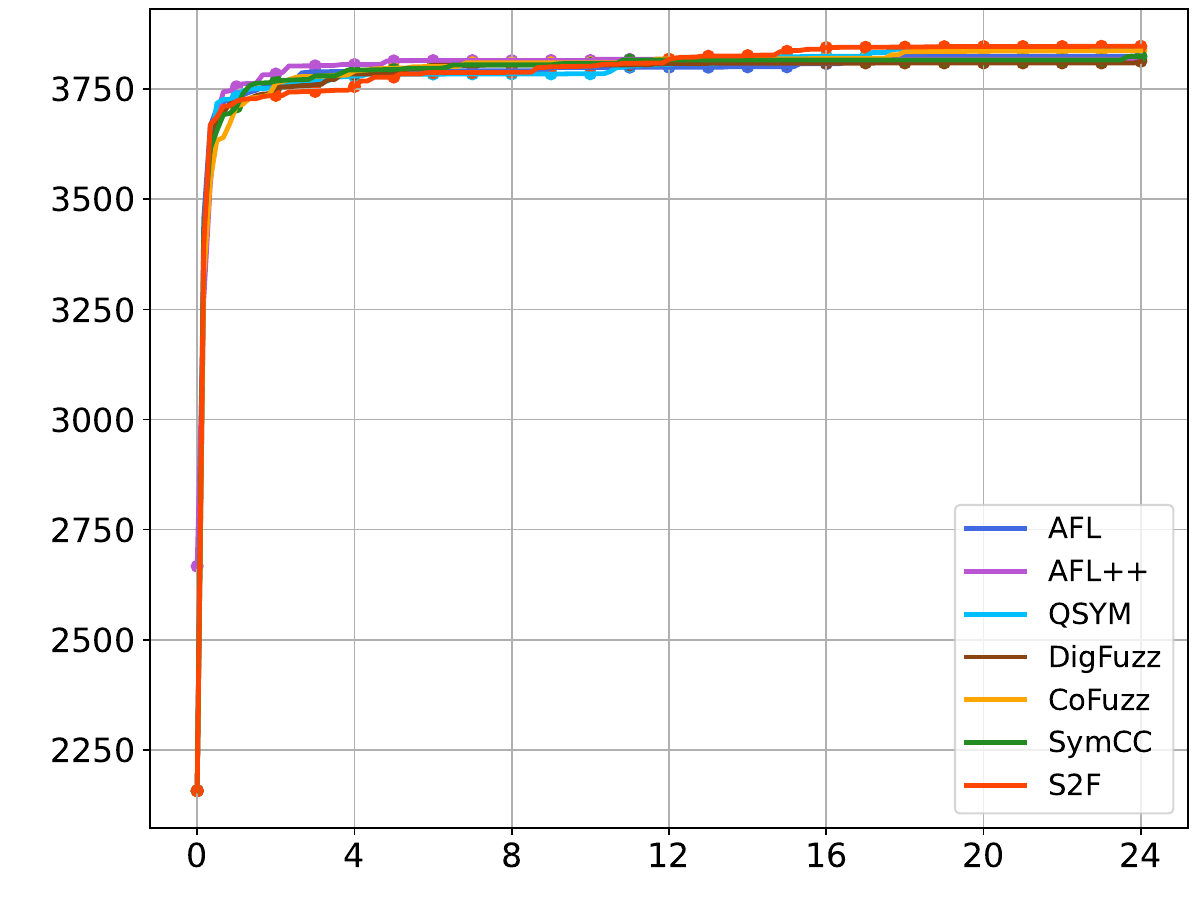}  
		\captionsetup{aboveskip=-1pt, belowskip=-1pt}
		\caption{libjpeg}  
		\label{fig:libjpeg}  
	\end{subfigure}
	\hfill  
	\begin{subfigure}[b]{0.195\textwidth}
		\centering  
		\includegraphics[width=\textwidth]{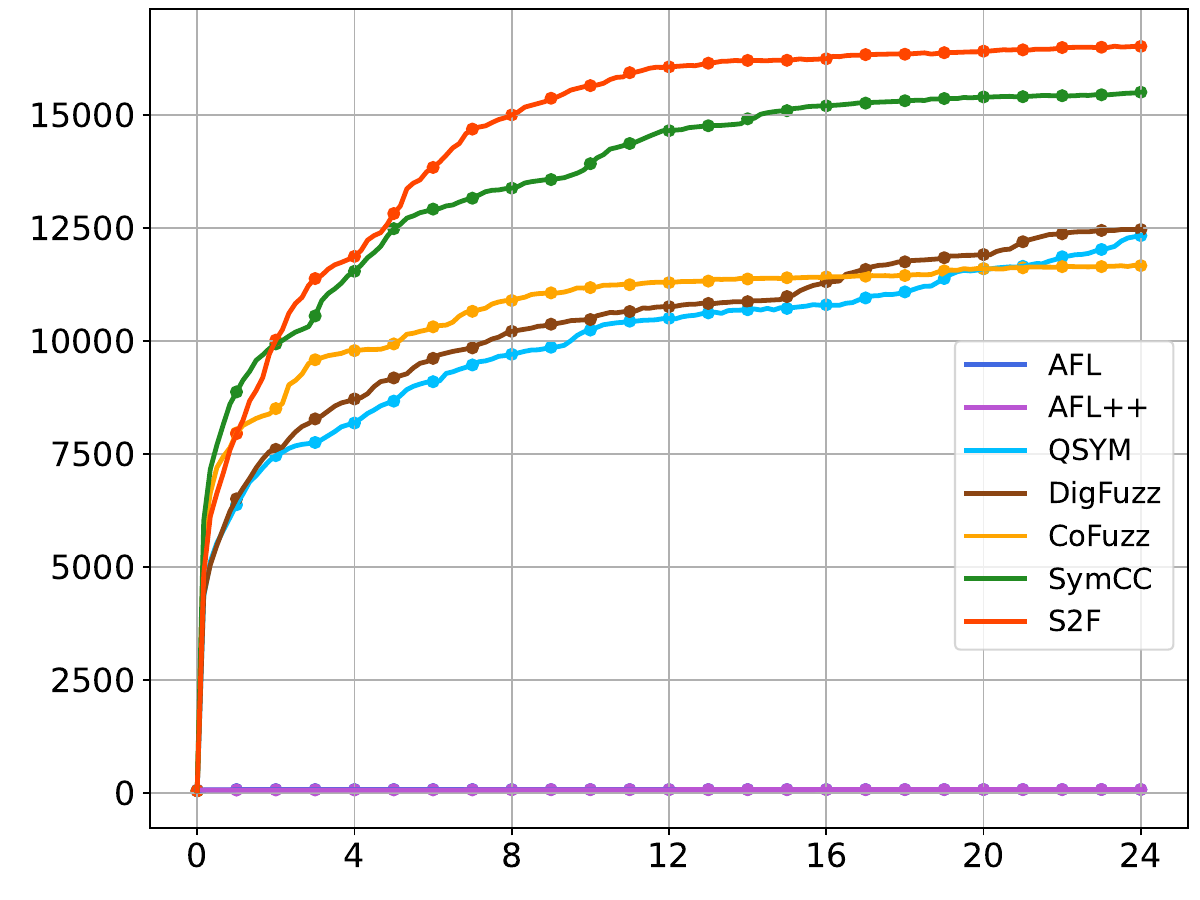}  
		\captionsetup{aboveskip=-1pt, belowskip=-1pt}
		\caption{tcpdump}  
		\label{fig:tcpdump}  
	\end{subfigure}
	\hfill
	\begin{subfigure}{0.195\textwidth}  
		\centering  
		\includegraphics[width=\textwidth]{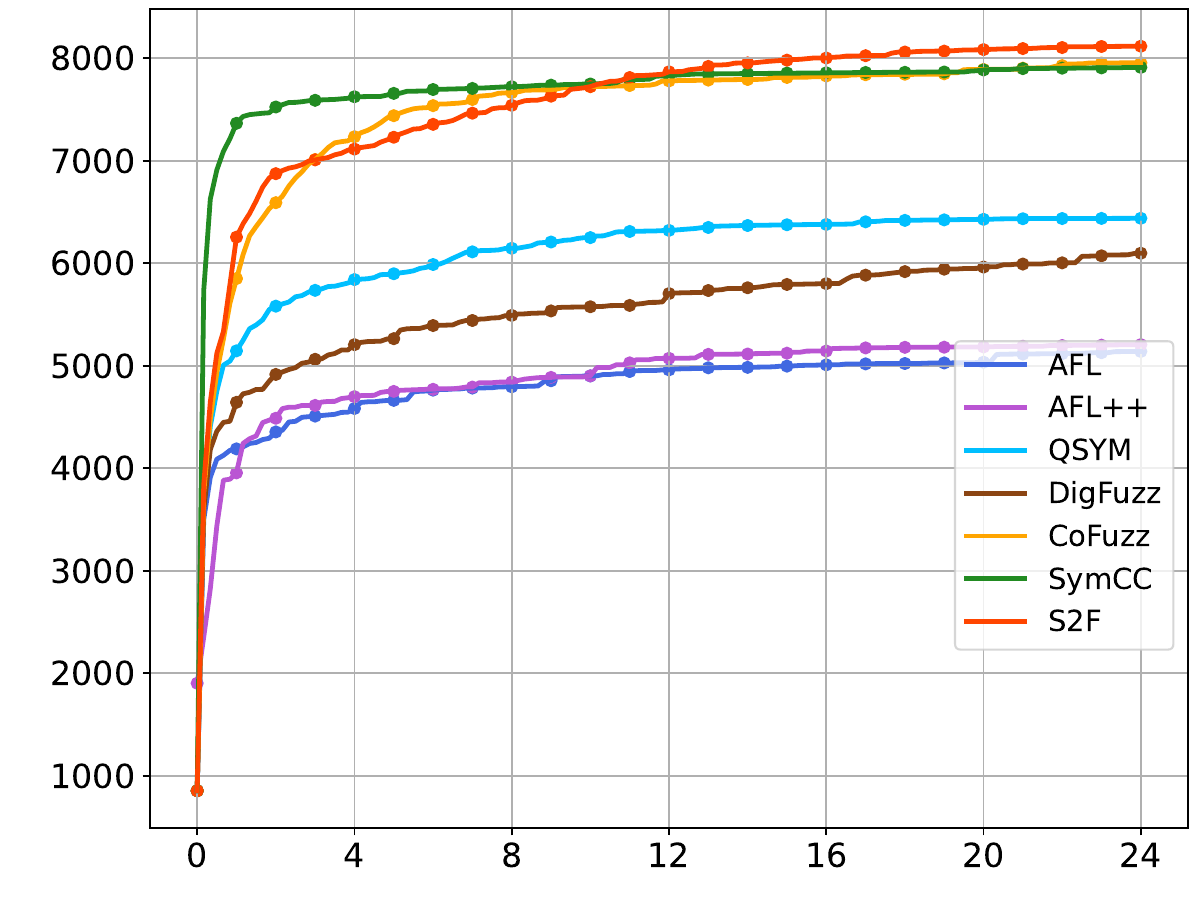}  
		\captionsetup{aboveskip=-1pt, belowskip=-1pt}
		\caption{libarchive}  
		\label{fig:libarchive}  
	\end{subfigure} 
	\hfill
	\begin{subfigure}{0.195\textwidth}  
		\centering  
		\includegraphics[width=\textwidth]{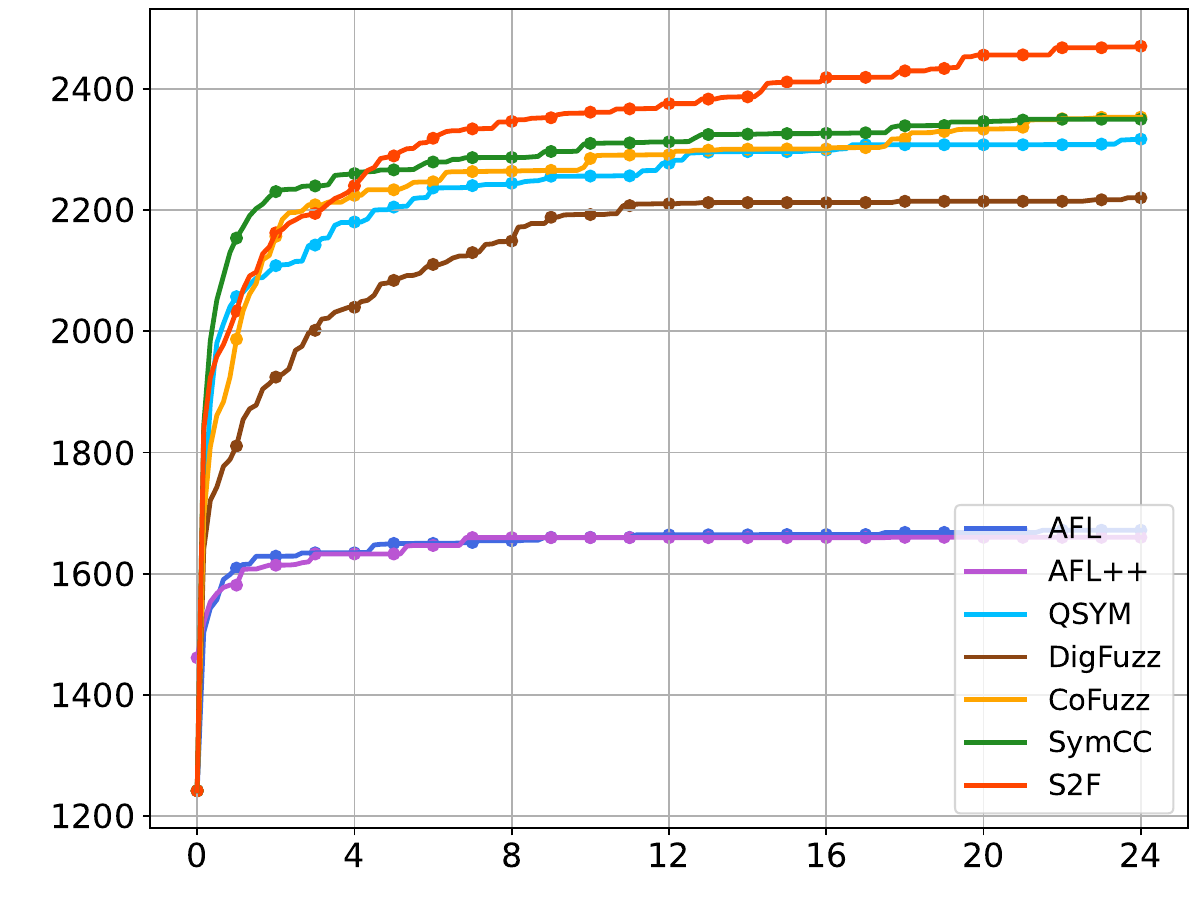}  
		\captionsetup{aboveskip=-1pt, belowskip=-1pt}
		\caption{pngimage}  
		\label{fig:pngimage}  
	\end{subfigure} 
	\hfill
	\begin{subfigure}{0.195\textwidth}  
		\centering  
		\includegraphics[width=\textwidth]{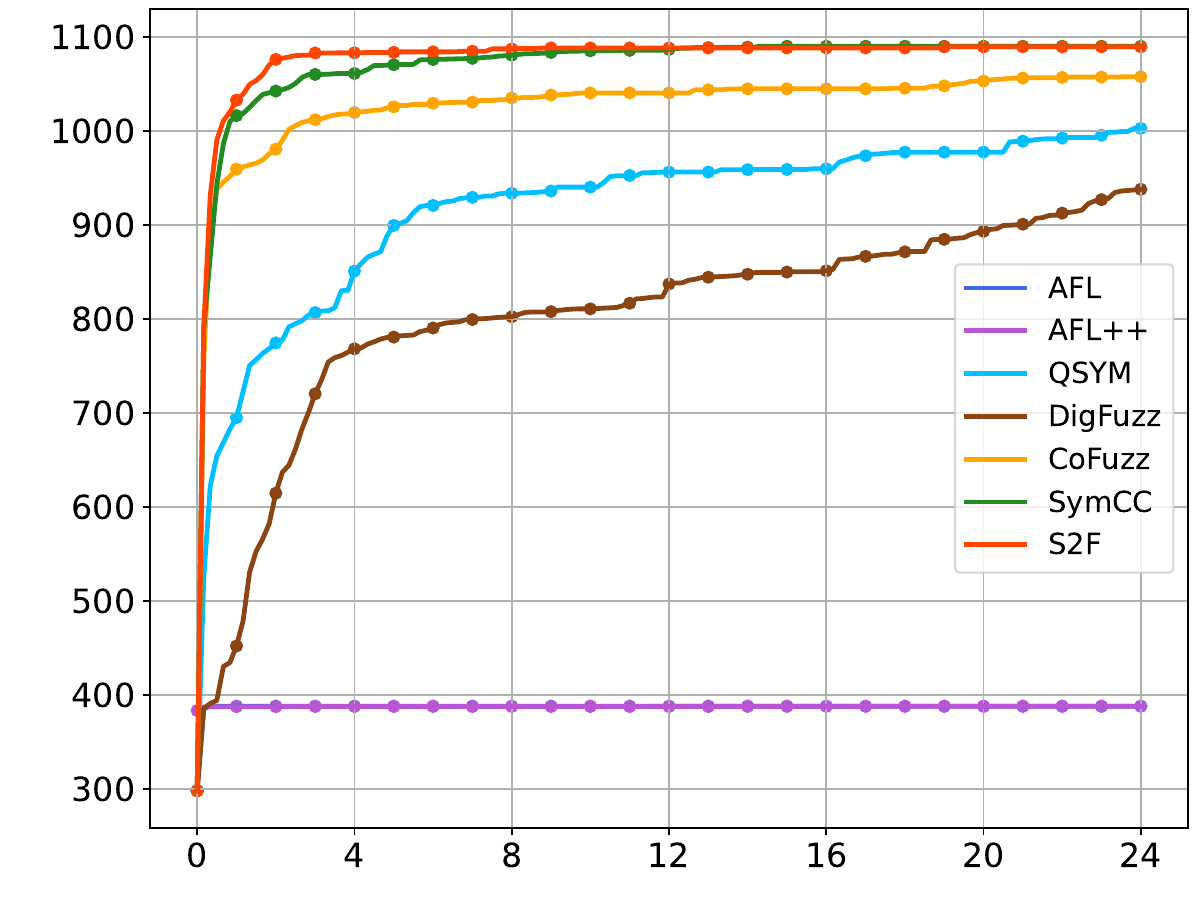}  
		\captionsetup{aboveskip=-1pt, belowskip=-1pt}
		\caption{jhead}  
		\label{fig:jhead}  
	\end{subfigure}
	\hfill
    \begin{subfigure}{0.195\textwidth}  
		\centering  
		\includegraphics[width=\textwidth]{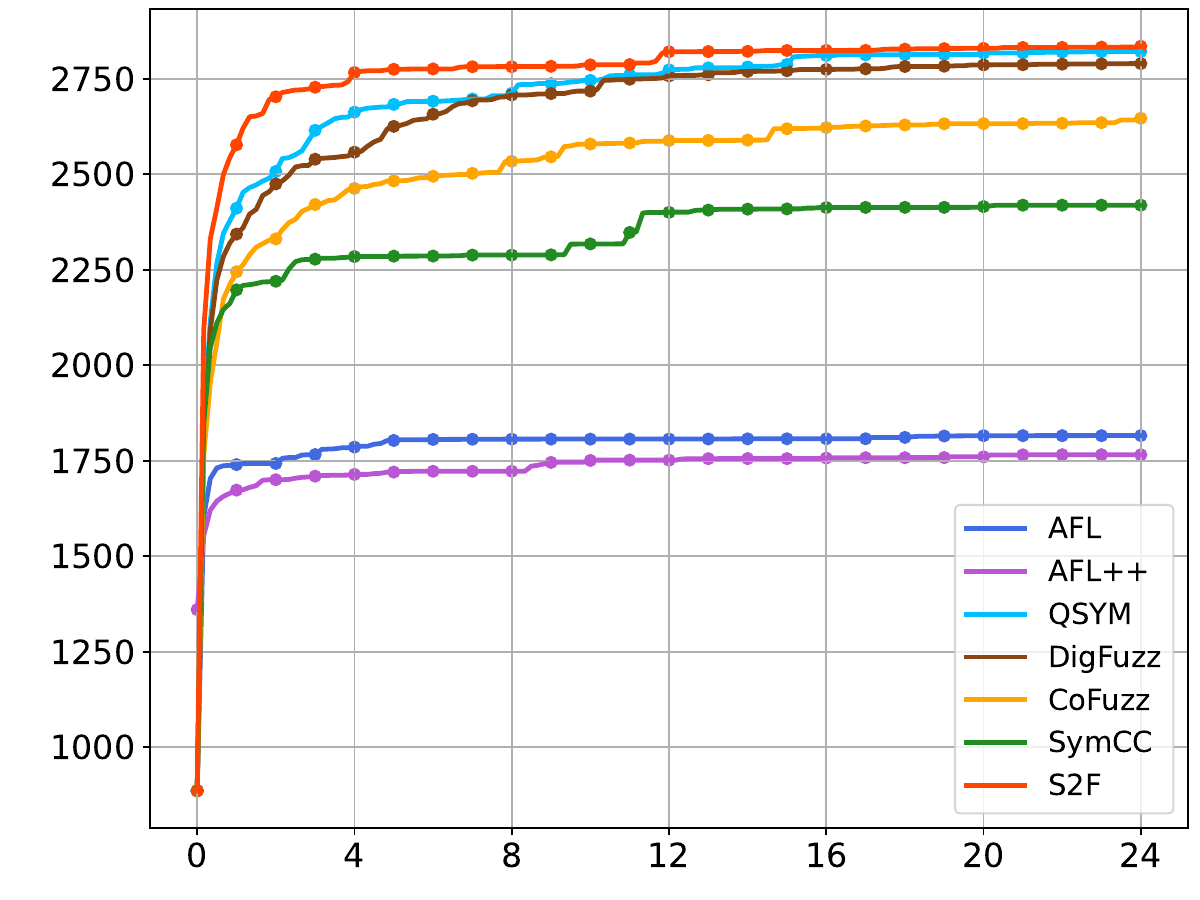}  
		\captionsetup{aboveskip=-1pt, belowskip=-1pt}
		\caption{pngfix}  
		\label{fig:pngfix}  
	\end{subfigure}
	\hfill
	\begin{subfigure}{0.195\textwidth}  
		\centering  
		\includegraphics[width=\textwidth]{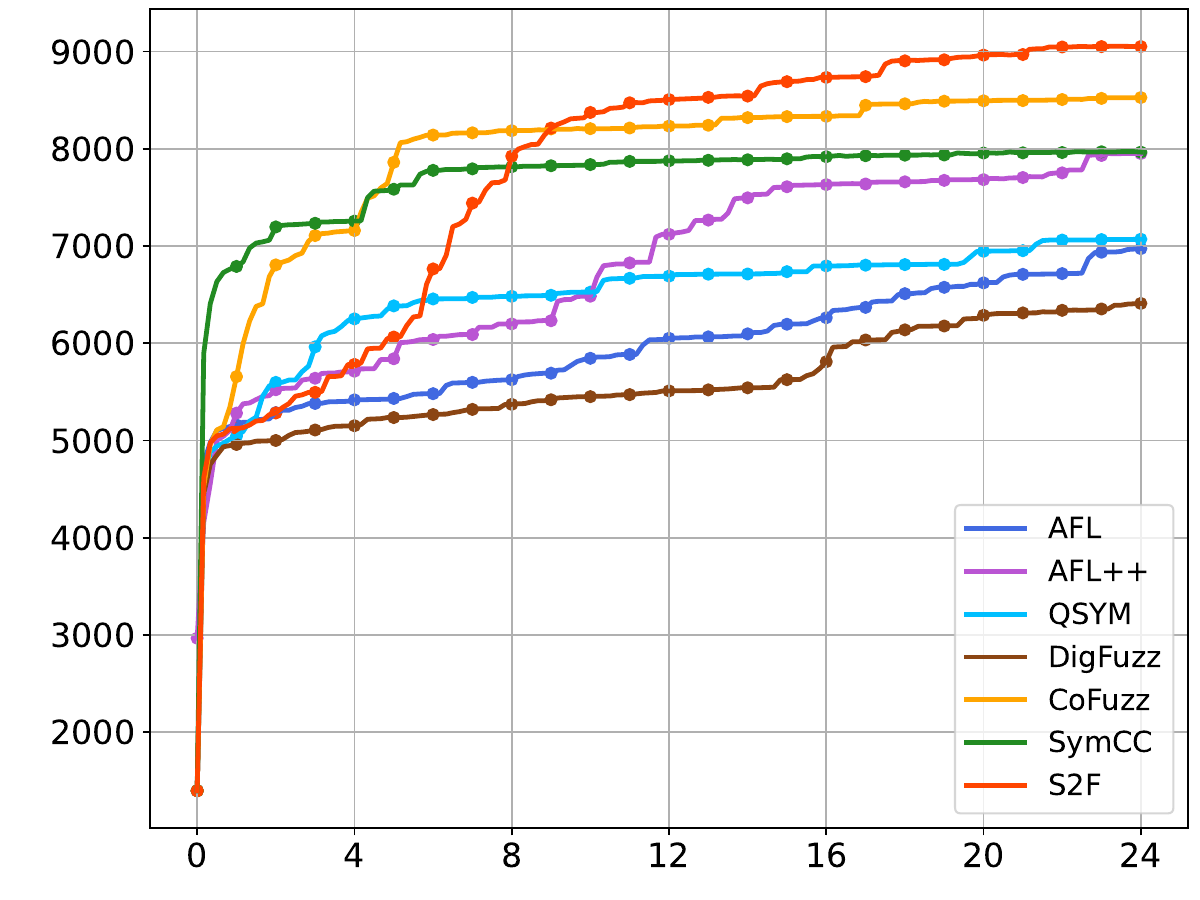}  
		\captionsetup{aboveskip=-1pt, belowskip=-1pt}
		\caption{libxml2}  
		\label{fig:libxml2}  
	\end{subfigure}
	\hfill 
	\begin{subfigure}{0.195\textwidth}  
		\centering  
		\includegraphics[width=\textwidth]{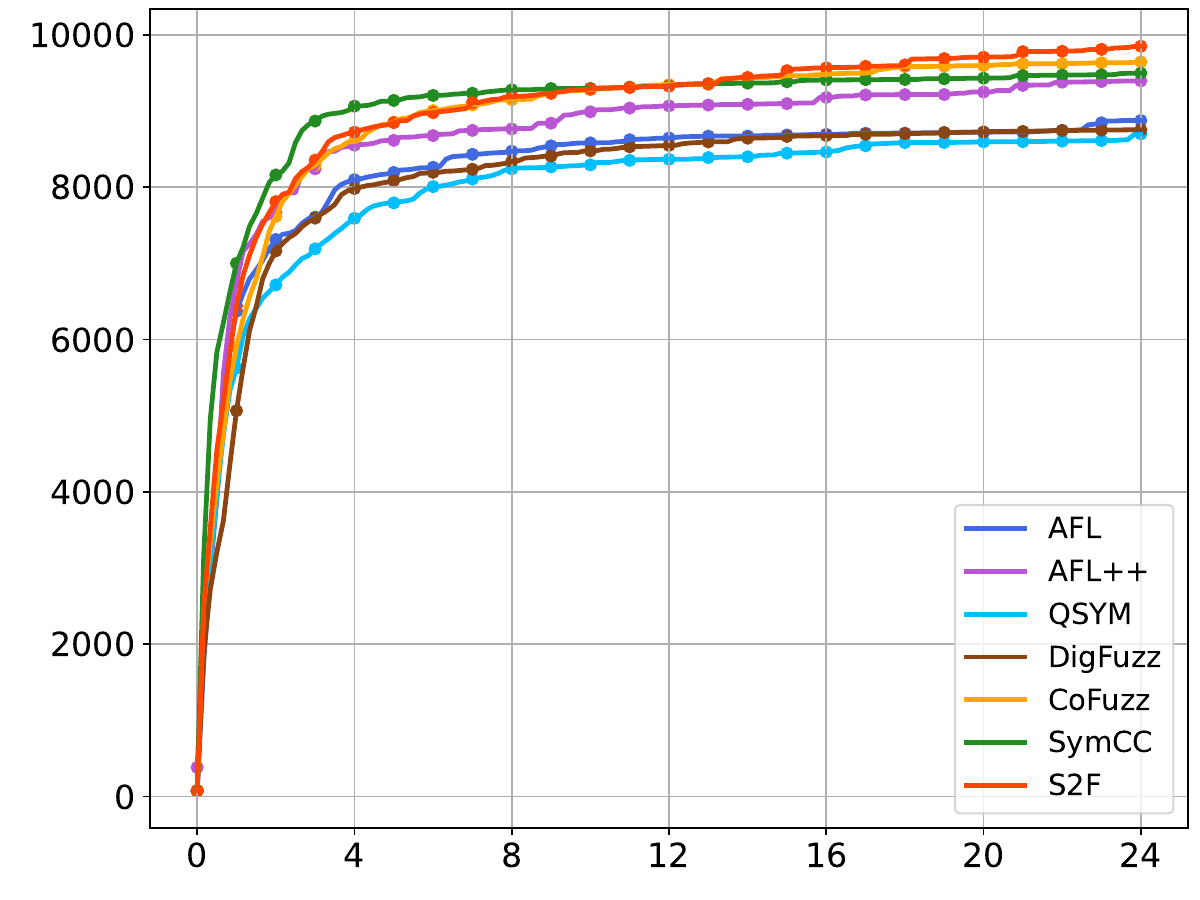}  
		\captionsetup{aboveskip=-1pt, belowskip=-1pt}
		\caption{readelf}  
		\label{fig:readelf}  
	\end{subfigure} 
	\hfill
    \begin{subfigure}{0.195\textwidth}  
		\centering  
		\includegraphics[width=\textwidth]{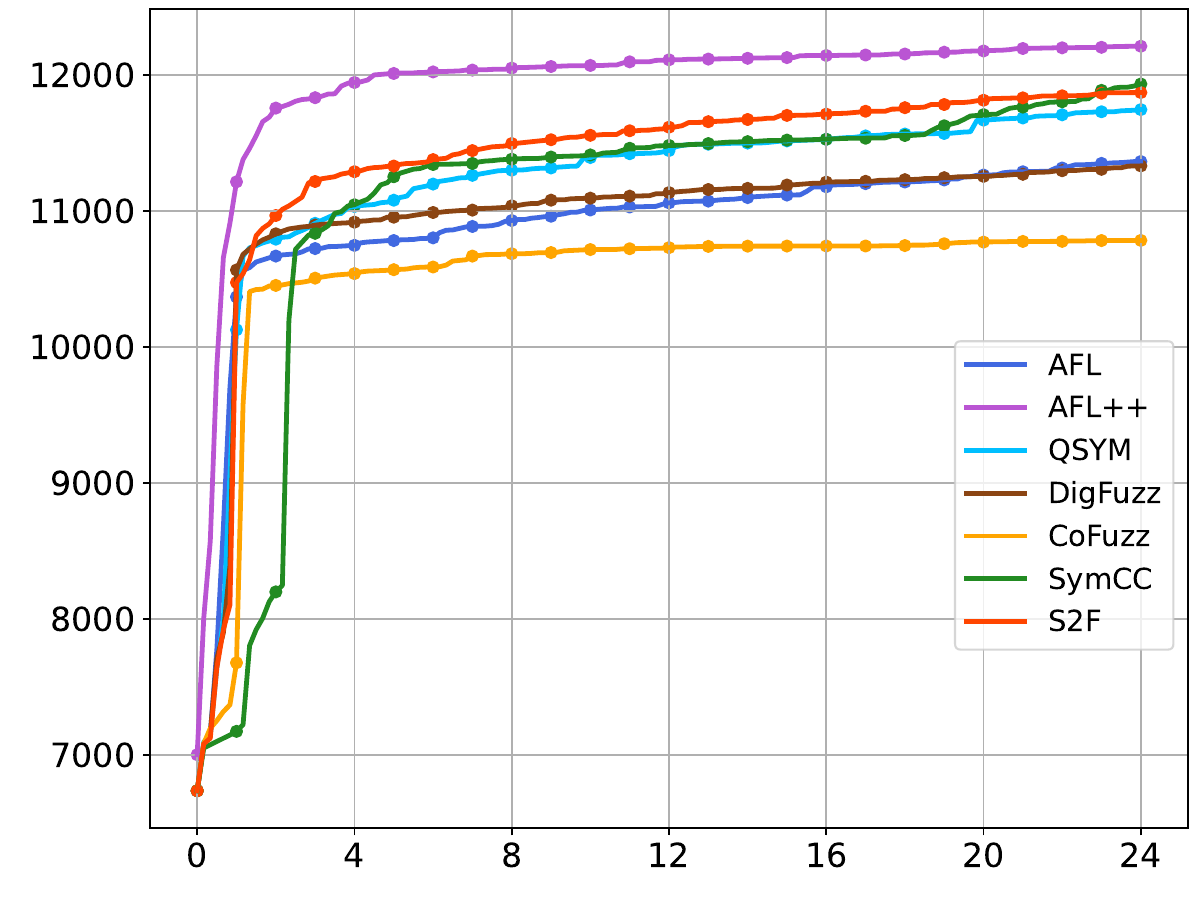}  
		\captionsetup{aboveskip=-1pt, belowskip=-1pt}
		\caption{openjpeg}  
		\label{fig:openjpeg}  
	\end{subfigure}
	\hfill 
	\begin{subfigure}{0.195\textwidth}  
		\centering  
		\includegraphics[width=\textwidth]{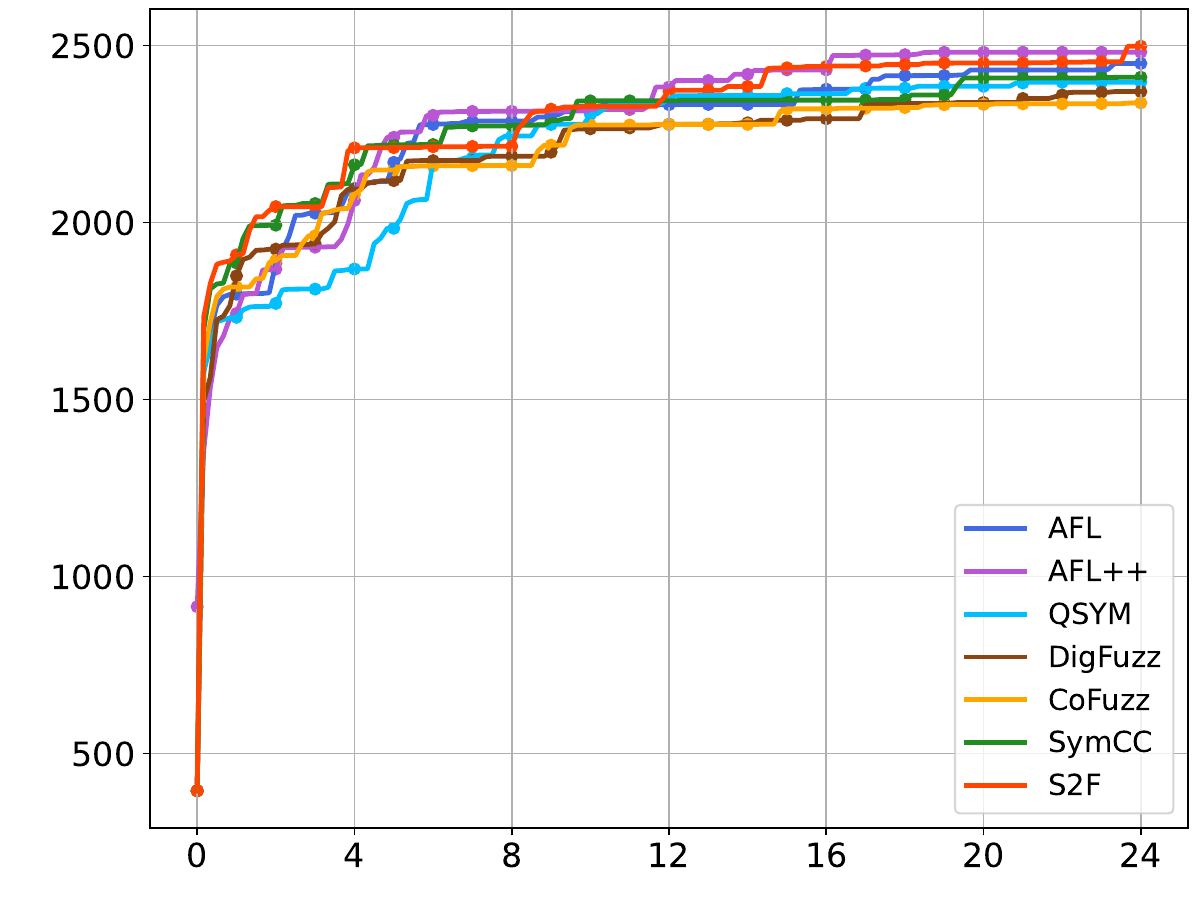}  
		\captionsetup{aboveskip=-1pt, belowskip=-1pt}
		\caption{gdk}  
		\label{fig:gdk}  
	\end{subfigure}
	\hfill
    \begin{subfigure}{0.195\textwidth}  
		\centering  
		\includegraphics[width=\textwidth]{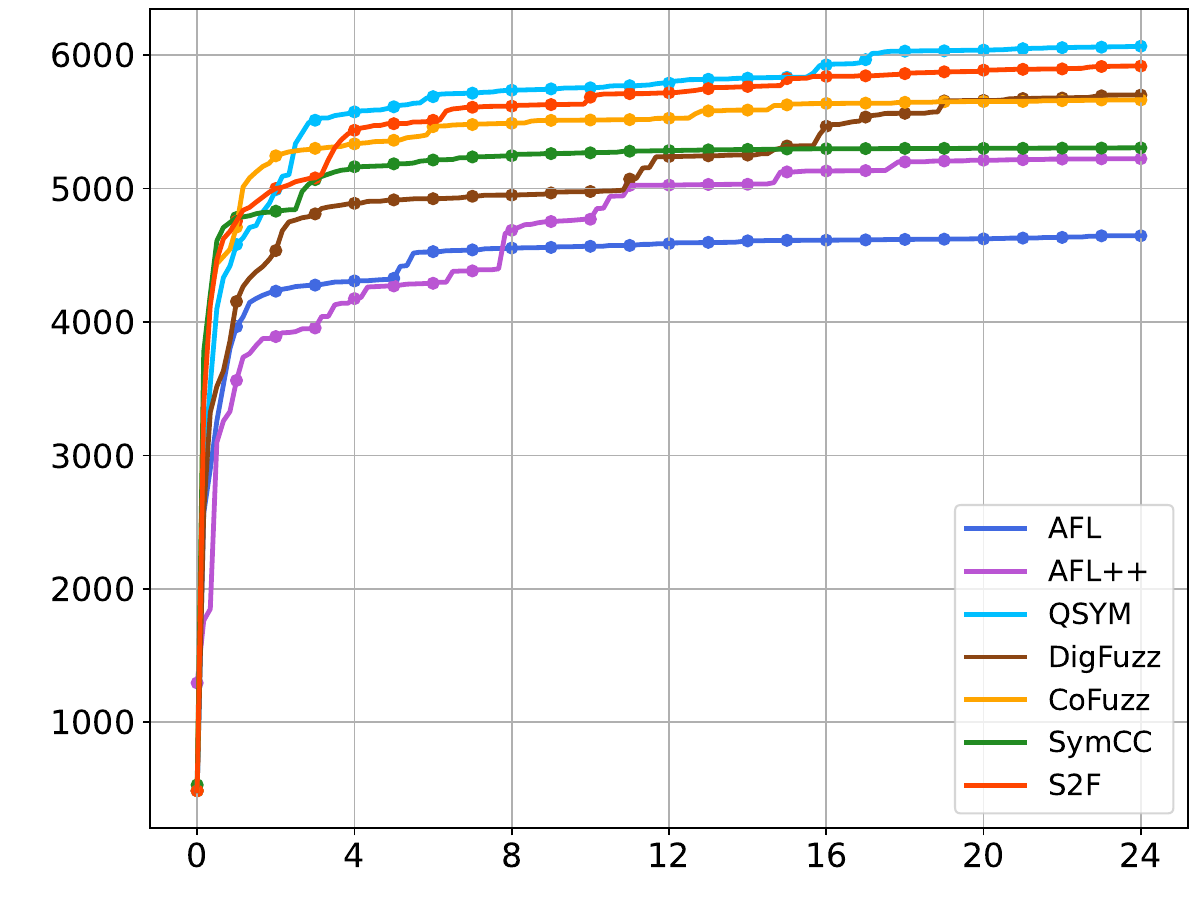}  
		\captionsetup{aboveskip=-1pt, belowskip=-1pt}
		\caption{nm}  
		\label{fig:nm}  
	\end{subfigure}
	\hfill
	\begin{subfigure}{0.195\textwidth}  
		\centering  
		\includegraphics[width=\textwidth]{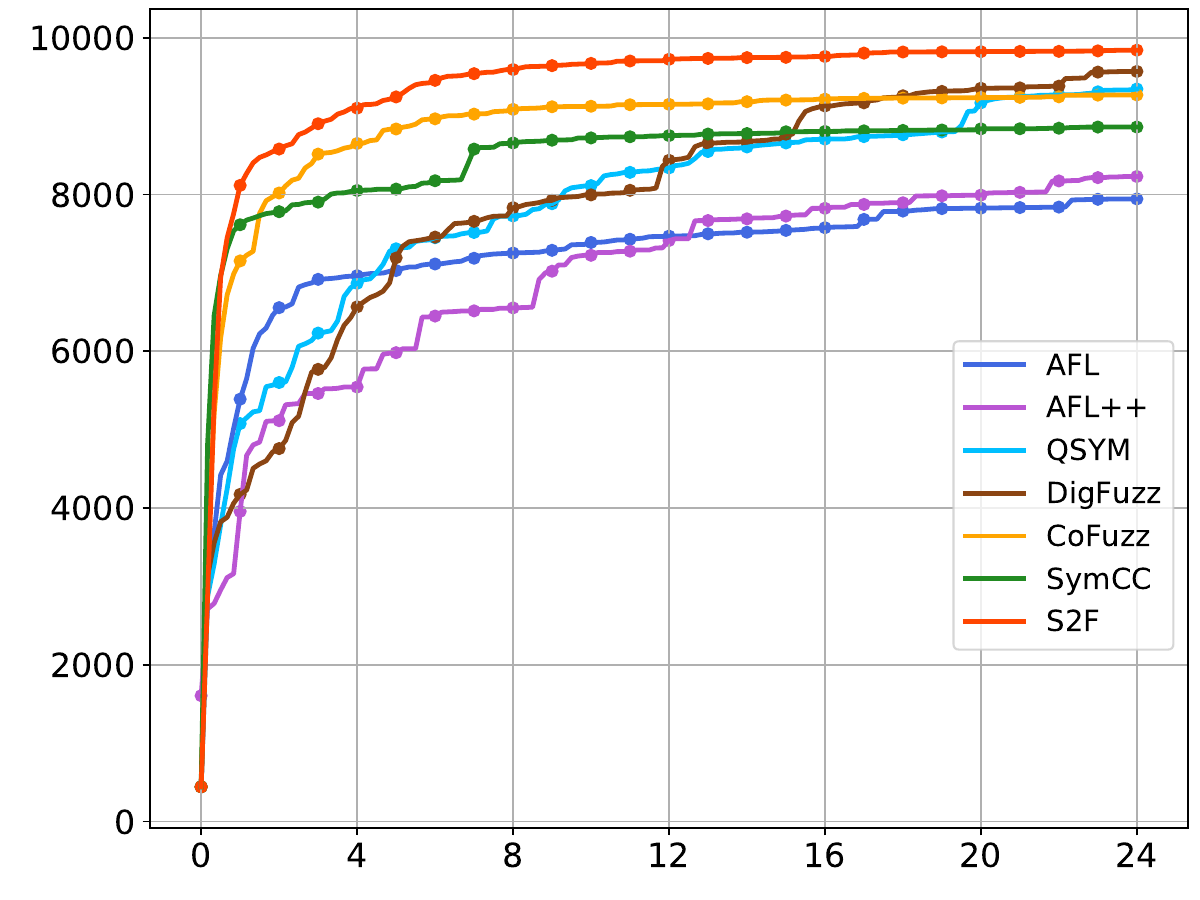}  
		\captionsetup{aboveskip=-1pt, belowskip=-1pt}
		\caption{strip}  
		\label{fig:strip}  
	\end{subfigure} 
	\hfill
	\begin{subfigure}{0.195\textwidth}  
		\centering  
		\includegraphics[width=\textwidth]{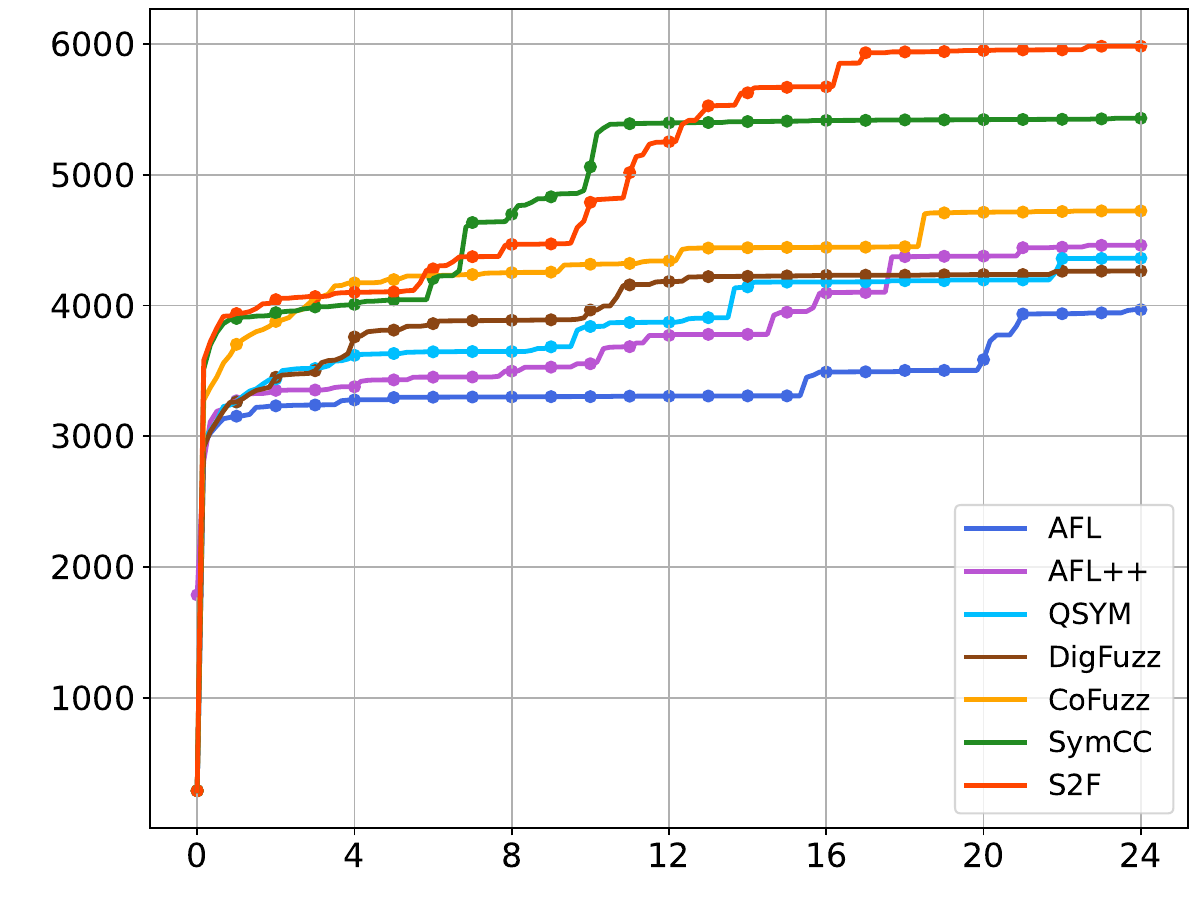}  
		\captionsetup{aboveskip=-1pt, belowskip=-1pt}
		\caption{imginfo}  
		\label{fig:imginfo}  
	\end{subfigure}
    \hfill  
    \begin{subfigure}{0.195\textwidth}  
		\centering  
		\includegraphics[width=\textwidth]{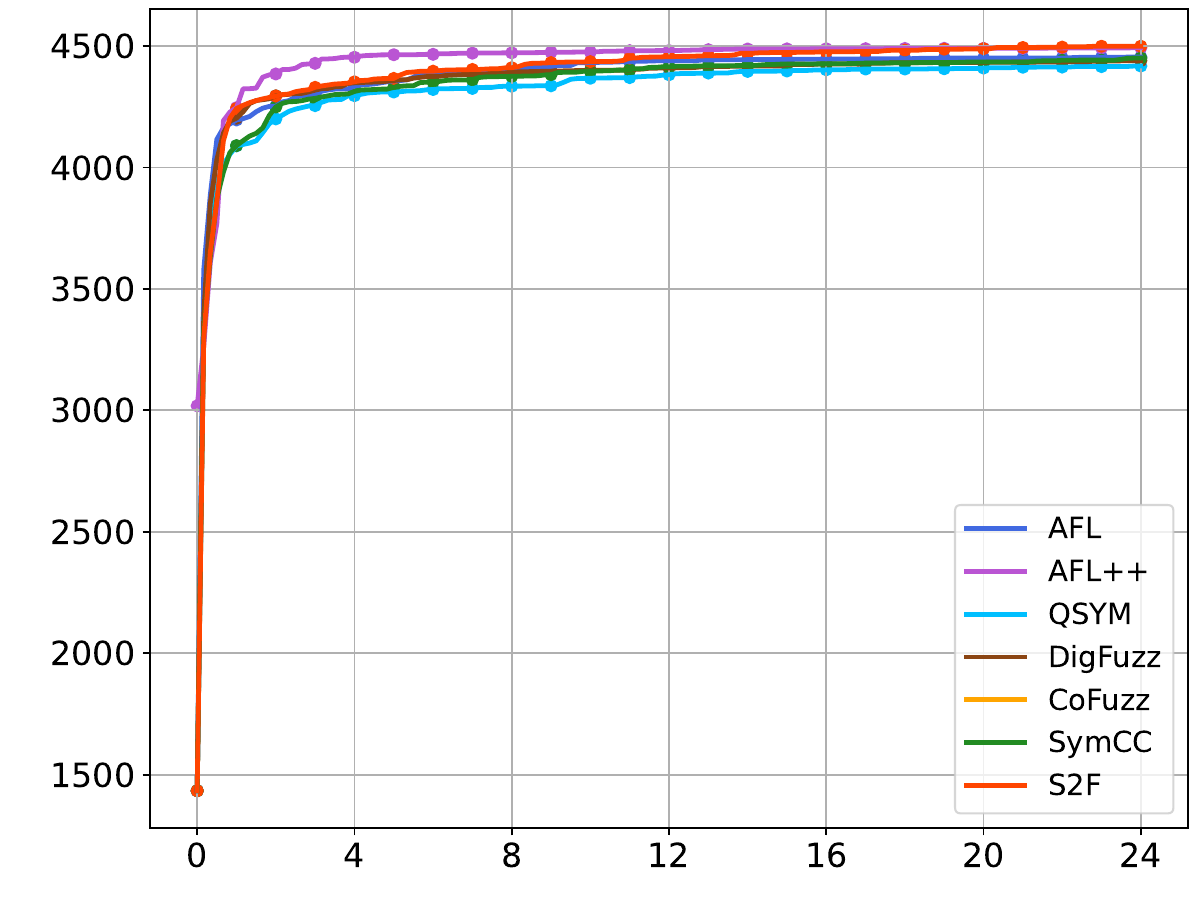}  
		\captionsetup{aboveskip=-1pt, belowskip=-1pt}
		\caption{cyclonedds}  
		\label{fig:cyclonedds}  
	\end{subfigure}  
	\caption{ Average edge coverage of 6 runs of 24 hours on 15 real-world programs. The X-axis is the running time in hours. The Y-axis is edge coverage. The higher the better.}
	\label{fig:images}
\end{figure*}

We selected the above tools for the following reasons.
AFL and AFL++ are widely used baselines in fuzzing research.
According to the results reported in \cite{Co_Fuzz}, CoFuzz and QSYM achieve the best and second-best performance when compared with Angora \cite{Angora}, Eclipser \cite{Eclipser}, Intriguer \cite{Intriguer}, DigFuzz \cite{digfuzz_on_qsym}, MEUZZ \cite{MEUZZ}, and PANGOLIN \cite{Pangolin}.
We also include SymCC, as it is generally regarded as more efficient than QSYM.
All the hybrid testing tools integrate AFL as the fuzzer.

We do not compare against PANGOLIN \cite{Pangolin} because it is not open source and focuses primarily on sampling techniques, which are orthogonal to our method for determining where to sample. 
We also exclude orthogonal approaches such as SymQEMU \cite{SymQemu}, SymFusion \cite{Sym_Fusion}, and SYMSAN \cite{SymSAN}, which aim to improve symbolic computation efficiency and adopt the same LOB strategy as SymCC.
Our method can be integrated into these tools as well.
Some other hybrid tools, such as \cite{rare}, which do not run fuzzing and symbolic execution in parallel, are also excluded.

As in prior work, each tool is executed with three parallel instances.
AFL and AFL++ use three fuzzing instances, whereas hybrid testing tools use two fuzzing instances and one symbolic execution instance.

\begin{table*}[t!]
    \setlength{\tabcolsep}{2.7pt}
    \caption{Average edge coverage (in percentage) of 6 runs of 24-hour evaluation. Each cell in the table contains the edge coverage (upper) and our improvement with respect to the tool (lower). The last two rows show our improvement with respect to each tool across all programs and for the 12 programs with more than 10K LOC. }
    \label{tab:edge coverage}     
    
    \centering
    \begin{minipage}{0.67\textwidth}
    \centering
    \begin{threeparttable}
    \begin{tabular}{l *{7}{r}}
        \toprule[0.4mm]
        \noalign{\smallskip}
        Program & AFL & AFL++ & QSYM &DigFuzz & CoFuzz &  SymCC  & \method\ (our)   \\\noalign{\smallskip}
        \hline\noalign{\smallskip}
                       objdump     & \makecell[r]{8652.67 \\[-1pt] \tiny{+18.90\%}} 
                        & \makecell[r]{9380.33 \\[-1pt] \tiny{+9.67\%}} 
                        & \makecell[r]{9887.17 \\[-1pt] \tiny{+4.05\%}}
                        & \makecell[r]{8895.33 \\\tiny{+15.65\%}} 
                         
                        & \makecell[r]{8737.33 \\\tiny{+17.74\%}} 
                        & \makecell[r]{9765.00 \\\tiny{+5.35\%}}  
                        & \makecell[r]{\textbf{10287.67}} \\
        \arrayrulecolor{gray!70}\hline
                libjpeg     & \makecell[r]{3825.00 \\\tiny{+0.59\%}} 
                        & \makecell[r]{3818.00 \\[-1pt] \tiny{+0.77\%}}
                        & \makecell[r]{3842.50 \\\tiny{+0.13\%}}
                        & \makecell[r]{3809.50 \\\tiny{+1.00\%}} 
                         
                        & \makecell[r]{3837.50 \\[-1pt] \tiny{+0.26\%}} 
                        & \makecell[r]{3824.17 \\[-1pt] \tiny{+0.61\%}} 
                        & \makecell[r]{\textbf{3847.50}} \\
        \arrayrulecolor{gray!70}\hline
            tcpdump     & \makecell[r]{81.00 \\[-1pt] \tiny{+20331.69\%}} 
                        & \makecell[r]{81.00 \\\tiny{+20331.69\%}}
                        & \makecell[r]{12381.33 \\[-1pt] \tiny{+33.67\%}}
                        & \makecell[r]{12494.17 \\[-1pt] \tiny{+32.46\%}} 
                        
                        & \makecell[r]{11685.17 \\[-1pt] \tiny{+41.63\%}}
                        & \makecell[r]{15512.50 \\\tiny{+6.69\%}}
                        & \makecell[r]{\textbf{16549.67}} \\
        \arrayrulecolor{gray!70}\hline
            libarchive  & \makecell[r]{5139.50 \\[-1pt] \tiny{+57.93\%}} 
                        & \makecell[r]{5206.17 \\[-1pt] \tiny{+55.91\%}}
                        & \makecell[r]{6438.33 \\[-1pt] \tiny{+26.07\%}}
                        & \makecell[r]{6097.83 \\[-1pt] \tiny{+33.11\%}}
                        
                        & \makecell[r]{7955.67\\[-1pt] \tiny{+2.03\%}}  
                        & \makecell[r]{7910.33 \\[-1pt] \tiny{+2.61\%}}
                        & \makecell[r]{\textbf{8116.83}} \\
        \hline
           pngimage      & \makecell[r]{1671.83 \\[-1pt] \tiny{+47.86\%}} 
                        & \makecell[r]{1660.17 \\[-1pt] \tiny{+48.90\%}}
                        & \makecell[r]{2317.00 \\[-1pt] \tiny{+6.69\%}}
                        & \makecell[r]{2220.33 \\[-1pt] \tiny{+11.33\%}} 
                         
                        & \makecell[r]{2353.33 \\[-1pt] \tiny{+5.04\%}} 
                        & \makecell[r]{2349.83 \\[-1pt] \tiny{+5.20\%}}  
                        & \makecell[r]{\textbf{2472.00}} \\
        \hline
            jhead       & \makecell[r]{388.00 \\[-1pt] \tiny{+180.71\%}} 
                        & \makecell[r]{387.83 \\[-1pt] \tiny{+180.84\%}}
                        & \makecell[r]{1002.50 \\[-1pt] \tiny{+8.65\%}}
                        & \makecell[r]{938.67 \\[-1pt] \tiny{+16.03\%}} 
                         
                        & \makecell[r]{1057.33 \\[-1pt] \tiny{+3.01\%}} 
                        & \makecell[r]{\textbf{1089.83} \\[-1pt] \tiny{-0.06\%}}  
                        & \makecell[r]{1089.17} \\
        \hline
            pngfix      & \makecell[r]{1816.00 \\[-1pt] \tiny{+56.13\%}} 
                        & \makecell[r]{1765.83 \\[-1pt] \tiny{+60.57\%}}
                        & \makecell[r]{2821.33 \\[-1pt] \tiny{+0.50\%}}
                        & \makecell[r]{2789.83 \\[-1pt] \tiny{+1.63\%}} 
                         
                        & \makecell[r]{2646.83 \\[-1pt] \tiny{+7.12\%}} 
                        & \makecell[r]{2419.17 \\[-1pt] \tiny{+17.20\%}}  
                        & \makecell[r]{\textbf{2835.33}} \\
         \hline                
            libxml2     & \makecell[r]{6971.33 \\[-1pt] \tiny{+29.85\%}} 
                        & \makecell[r]{7951.00 \\[-1pt] \tiny{+13.85\%}}
                        & \makecell[r]{7069.00 \\[-1pt] \tiny{+28.06\%}}
                        & \makecell[r]{6408.83 \\[-1pt] \tiny{+41.25\%}} 
                         
                        & \makecell[r]{8521.17 \\[-1pt] \tiny{+6.24\%}} 
                        & \makecell[r]{7967.17 \\[-1pt] \tiny{+13.62\%}}  
                        & \makecell[r]{\textbf{9052.50}} \\                       
    
        \hline  
    
            readelf     & \makecell[r]{8873.67 \\[-1pt] \tiny{+10.99\%}} 
                        & \makecell[r]{9393.83 \\[-1pt] \tiny{+4.84\%}}
                        & \makecell[r]{8698.00 \\[-1pt] \tiny{+13.23\%}}
                        & \makecell[r]{8752.67 \\[-1pt] \tiny{+12.52\%}} 
                         
                        & \makecell[r]{9643.50 \\[-1pt] \tiny{+2.13\%}} 
                        & \makecell[r]{9494.83 \\[-1pt] \tiny{+3.72\%}} 
                        & \makecell[r]{\textbf{9848.50}} \\
        \hline
        openjpeg    & \makecell[r]{11364.50  \\[-1pt]\tiny{+4.47\%}} 
                    & \makecell[r]{\textbf{12212.33} \\[-1pt] \tiny{-2.79\%}} 
                    & \makecell[r]{11746.33 \\[-1pt] \tiny{+1.07\%}}
                    & \makecell[r]{11332.33 \\[-1pt] \tiny{+4.76\%}} 
                    
                    & \makecell[r]{10784.33 \\[-1pt] \tiny{+10.09\%}}
                    & \makecell[r]{11934.67 \\[-1pt] \tiny{-0.52\%}}
                    & \makecell[r]{11872.17} \\                
        \hline
            gdk         & \makecell[r]{2449.50 \\[-1pt] \tiny{+2.05\%}} 
                        & \makecell[r]{2481.33 \\[-1pt] \tiny{+0.74\%}}
                        & \makecell[r]{2396.50 \\[-1pt] \tiny{+4.31\%}}
                        & \makecell[r]{2370.00 \\[-1pt] \tiny{+5.47\%}} 
                         
                        & \makecell[r]{2338.17 \\[-1pt] \tiny{+6.91\%}} 
                        & \makecell[r]{2410.67 \\[-1pt] \tiny{+3.69\%}}  
                        & \makecell[r]{\textbf{2499.67}} \\
        \hline
        nm          & \makecell[r]{4646.00 \\[-1pt] \tiny{+27.38\%}} 
                    & \makecell[r]{5223.67 \\[-1pt] \tiny{+13.29\%}}
                    & \makecell[r]{\textbf{6066.67} \\[-1pt] \tiny{-2.45\%}}
                    & \makecell[r]{5700.67 \\[-1pt] \tiny{+3.81\%}} 
                        
                    & \makecell[r]{5664.00 \\[-1pt] \tiny{+4.48\%}} 
                    & \makecell[r]{5305.17 \\[-1pt] \tiny{+11.55\%}} 
                    & \makecell[r]{5918.00} \\ 
        \hline
        strip        & \makecell[r]{7943.33 \\[-1pt] \tiny{+23.90\%}} 
                    & \makecell[r]{8229.67 \\[-1pt] \tiny{+19.59\%}}
                    & \makecell[r]{9346.33 \\[-1pt] \tiny{+5.30\%}}
                    & \makecell[r]{9570.00 \\[-1pt] \tiny{+2.84\%}} 
                        
                    & \makecell[r]{9270.00 \\[-1pt] \tiny{+6.17\%}} 
                    & \makecell[r]{8861.00 \\[-1pt] \tiny{+11.07\%}}  
                    & \makecell[r]{\textbf{9841.50}} \\                           
        \hline
    
            imginfo     & \makecell[r]{3970.33 \\[-1pt] \tiny{+50.70\%}} 
                        & \makecell[r]{4461.00 \\[-1pt] \tiny{+34.13\%}} 
                        & \makecell[r]{4362.17 \\[-1pt] \tiny{+37.16\%}}
                        & \makecell[r]{4264.67 \\[-1pt] \tiny{+40.30\%}} 
                         
                        & \makecell[r]{4724.00 \\[-1pt] \tiny{+26.66\%}} 
                        & \makecell[r]{5432.67 \\[-1pt] \tiny{+10.14\%}}  
                        & \makecell[r]{\textbf{5983.33}} \\
        \hline
            cyclonedds   & \makecell[r]{4453.50 \\[-1pt] \tiny{+1.02\%}} 
                        & \makecell[r]{4491.33 \\[-1pt] \tiny{+0.17\%}}
                        & \makecell[r]{4417.33 \\[-1pt] \tiny{+1.85\%}}
                        & \makecell[r]{4439.17 \\[-1pt] \tiny{+1.34\%}} 
                            
                        & \makecell[r]{N}  
                        & \makecell[r]{4446.67 \\[-1pt] \tiny{+1.17\%}} 
                        & \makecell[r]{\textbf{4498.83}} \\
            \arrayrulecolor{black}\hline
            \makecell[l]{\scriptsize{average} \\[-1pt] \scriptsize{improvement}}
            & \makecell[r]{+36.60\%}
            & \makecell[r]{+31.46\%}
            & \makecell[r]{+11.22\%}
            & \makecell[r]{+14.90\%}
            & \makecell[r]{+9.96\%}
            & \makecell[r]{+6.14\%}   
            \\
            \arrayrulecolor{black}\hline
            \makecell[l]{\scriptsize{average} \scriptsize{improvement}\\[-1pt] \scriptsize{($>$10KLOC)}}
            & \makecell[r]{+29.88\%}
            & \makecell[r]{+23.52\%}
            & \makecell[r]{+12.79\%}
            & \makecell[r]{+16.72\%}
            & \makecell[r]{+10.80\%}
            & \makecell[r]{+7.27\%}   
            \\
            \bottomrule[0.4mm]
        \end{tabular}
        \end{threeparttable}  
    \end{minipage}%
    \begin{minipage}{0.33\textwidth} 
    \centering
    \includegraphics[width=\linewidth]{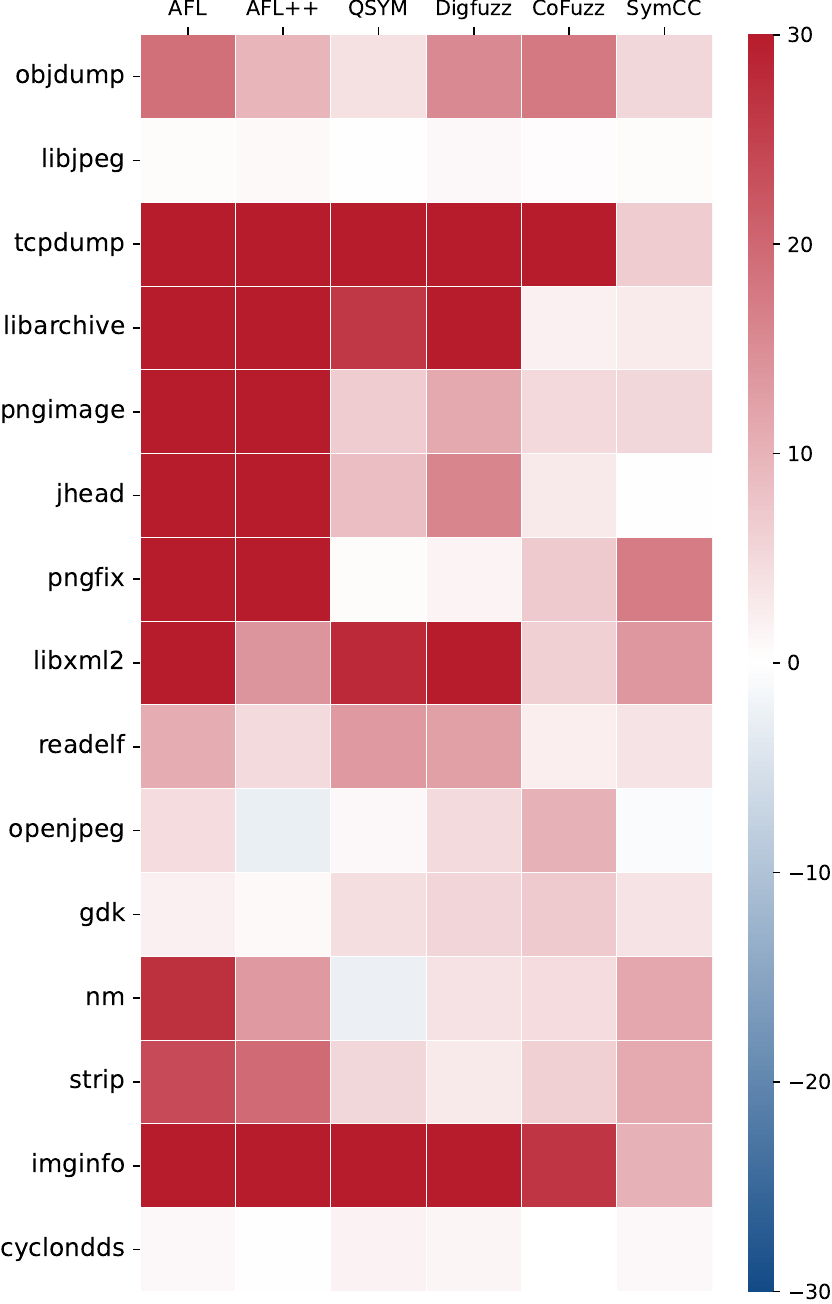}
    \captionof{figure}{{Percentage of coverage improvement by \method\ relative to the baseline tools. Red indicates that \method\ attains higher coverage, whereas blue indicates lower coverage compared to baselines.}}
    \label{fig:improve}
    \end{minipage}
    \end{table*}

 \textbf{Benchmark}.
Following prior work on hybrid testing and fuzzing (SymCC \cite{SymCC}, PANGOLIN \cite{Pangolin}, CoFuzz \cite{Co_Fuzz}), we construct a benchmark of 15 real-world programs (see Table~\ref{tab:2}). These programs are widely used in related research \cite{Pangolin, QSYM, SymCC, Co_Fuzz, SymSAN}. Programs that contain inline assembly or LLVM intrinsic functions (\textit{e.g.}, \textit{lame} and \textit{file}) are excluded, as these features are not supported in the current IR-based symbolic execution engine yet. We adopt the same initial seeds as prior work.

All experiments are conducted on servers with 32 cores (2.40 GHz
Intel Xeon Silver 4314), 128 GB of memory, and Ubuntu 20.04 LTS.
To alleviate the fluctuation caused by randomness,
each configuration is run 6 times to compute the average result. 
In total, the experiments consumed 97,200 CPU hours (4,050 CPU days) of computation.

\subsection{Coverage}
Fig.~\ref{fig:images} shows the edge-coverage curves over 24 hours, and the corresponding numerical results are provided in Table~\ref{tab:edge coverage}.
Fig.~\ref{fig:improve} reports the percentage of coverage improvement achieved by \method\ relative to the baseline tools.
Specifically, \method\ improves edge coverage over AFL, AFL++, QSYM, DigFuzz, CoFuzz, and SymCC by 36.60\%\footnote{We exclude the result of \textit{tcpdump} when computing the average improvement over AFL and AFL++, as including it would increase the averages to 1389.61\% and 1384.81\%, respectively.}, 31.46\%, 11.22\%, 14.90\%, 9.96\%, and 6.14\%, respectively.
Across the 15 programs, \method\ achieves the best results on 12 programs. AFL++, QSYM, and SymCC each achieve one win on \emph{openjpeg}, \emph{nm}, and \emph{jhead}, respectively, while the remaining tools do not win on any program.

Among the 15 programs, \emph{tcpdump} and \emph{jhead} are the two cases where hybrid testing achieves substantially higher coverage (over 20,331\% and 180\%) than pure fuzzing (AFL and AFL++). We manually inspected their code and found that both programs contain complicated syntax checking. Most mutations generated by fuzzing fail to pass these checks. In contrast, symbolic execution can generate valid inputs and cover subsequent program parts.

Specifically, although the number of lines of code is not always proportional to a program’s complexity, it can still reflect the difficulty of program testing.
We collect results on 12 programs exceeding 10K LOC. On these programs , \method\ outperforms AFL, AFL++, QSYM, DigFuzz, CoFuzz, and SymCC by  29.88\%, 23.52\%, 12.79\%, 16.72\%, 10.80\%, 7.27\%, respectively. 
Among these tools, \method\ surpasses the second-best tool, SymCC, by a larger margin than observed across all programs.
Notably, \method\ outperforms the two fuzzing tools, AFL and AFL++, by 29.88\% and 23.52\%, which are smaller than the improvements observed across all programs. This is because \emph{jhead}, on which hybrid testing significantly outperforms fuzzing, is excluded due to its small size (2K LOC).

For \emph{openjpeg}, AFL++ achieves the best result. When integrating with AFL++, \method\ can beat all other tools too (see Table \ref{tab:s2fafl++} discussed later).

These results indicate that our tool demonstrates greater advantages for larger programs. We attribute this improvement to multiple factors. One of the reasons is that existing hybrid testing tools use bitmaps to track edge coverage. For larger programs, hash collisions are more severe, making the LOB strategy prune more branches.

\subsection{Unique Crashes}
Table \ref{tab:Unique crashes} reports the number of unique crashes discovered by each tool. Following prior work \cite{unifuzz}, we filter out redundant crashes using the top three call stack frames. We conduct six independent 24-hour experiments and remove duplicate crashes across runs to calculate the total number of unique crashes. In total, \method\ identifies the highest count, 57 unique crashes, followed by AFL++ with 43.

Among these programs, AFL and AFL++ find more than 20 unique crashes in \emph{cyclonedds}, making fuzzing tools (AFL and AFL++) better than hybrid testing tools (except for ours) in total. Except for \emph{cyclonedds}, hybrid testing tools outperform fuzzing across the remaining programs. 

Overall, \method\ discovers more unique crashes than AFL, AFL++, QSYM, DigFuzz, CoFuzz, and SymCC by 39.0\%, 32.6\%, 58.3\%, 42.5\%, 72.7\%, and 58.3\%, respectively. 
These results demonstrate the superiority of our tool in discovering crashes.
\begin{table}[t]
    \center
    \caption{ {Unique crashes found in 6 runs of 24-hour evaluation.}}
    \label{tab:Unique crashes}       
     
            \label{Unique crashes of 6 runs of 24-hour evaluation. The higher the better.}
            \begin{tabular}{l C{0.5cm} C{0.5cm} C{0.5cm} C{0.5cm} C{0.6cm} C{0.5cm} C{1cm} } 
                \toprule[0.4mm]
                \scriptsize{Program} & \scriptsize{AFL}  & \scriptsize{AFL++} & \scriptsize{QSYM} & \scriptsize{DigFuzz}  & \scriptsize{CoFuzz} & \scriptsize{SymCC}  & \scriptsize{\method\ (our)}\\
                \noalign{\smallskip}\hline\noalign{\smallskip}
                objdump &8 &7 &5 &11 &8 &7  &13\\
                libxml2  & 0 & 0 & 1 & 0 & 1 & 0  & 0\\
                readelf  & 0 & 0 & 0 & 0 & 0 & 0 & 1\\
                gdk      & 9 & 9  & 9 & 9 & 10 &10&10\\
                strip &4& 3&5&5&12&6&11\\
                imginfo &0 &0 &2 & 1 &2 & 1   &4\\
                cyclonedds &20 & 24&14&14& N&12&18\\
                \noalign{\smallskip}\hline
                \noalign{\smallskip}
                total & 41  & 43 & 36 & 40 & 33 & 36  & \textbf{57}\\ 
                \hline\noalign{\smallskip}
                \scriptsize{improvement} &\tiny{+}\scriptsize{39.0\%} &\tiny{+}\scriptsize{32.6\%} &\tiny{+}\scriptsize{58.3\%} &\tiny{+}\scriptsize{42.5\%}  &\scriptsize{+72.7\%} &\tiny{+}\scriptsize{58.3\%}\\
                \bottomrule[0.4mm]
            \end{tabular}
   
    \end{table}

\textbf{0-day Vulnerabilities}. Table \ref{tab:bugs detected} lists the bugs discovered by \method\ during all experiments that were not found by other tools.
We check all the crashes manually. If the latest version of the program in the benchmark does not crash with the same input, we mark the crash as patched in Table \ref{tab:bugs detected}. 
Among the 17 crashes, 8 are known CVE vulnerabilities.
\method\ discovers three new vulnerabilities in \emph{cyclonedds}.
These results highlight the advantage of \method\ in uncovering vulnerabilities. 

Overall, we can draw the following conclusion.
\begin{mdframed}[
    skipabove=2pt,
    innertopmargin=2pt,
    innerbottommargin=2pt,
    linewidth=0.4pt
]
\textbf{Answer to Q1}: \method\ outperforms the state-of-the-art hybrid testing tools in terms of both coverage and discovered crashes. Therefore, our method can improve the testing efficiency of hybrid testing.
\end{mdframed}

\subsection{Sleeping Time}
The last column of Table \ref{tab:sleep} shows the average sleeping time of symbolic execution in our tool. On average, the symbolic executor in \method\ only spends 12.88\% of the running time waiting for AFL's seeds, which is significantly less than SymCC (56.42\%) and CoFuzz (37.61\%). Therefore, our method can reduce the sleeping time of symbolic execution effectively.

Specifically, the second-best tool, SymCC, sleeps for 11 programs in Table \ref{tab:sleep}. For these programs, \method\ only sleeps 17.56\% on average, whereas SymCC and CoFuzz spend 76.94\% and 50.83\% of time sleeping. \method\ outperforms AFL, AFL++, QSYM, DigFuzz, CoFuzz, and SymCC by 46.46\%, 41.32\%, 14.34\%, 18.12\%, 10.52\%, 6.87\%, respectively.

In summary, we can obtain the following conclusion.
\begin{mdframed}[
skipabove=2pt,
    innertopmargin=2pt,
    innerbottommargin=2pt,
  ]
	\textbf{\textbf{Answer to Q2}}: Our method can reduce the sleeping time of symbolic execution in hybrid testing significantly.
\end{mdframed}

\subsection{Ablation Study}
\begin{table}[t]
\caption{ Average edge coverage of 6 runs of 24-hour evaluation. The last row shows the improvement of all configurations with respect to the first configuration.}
\label{tab:ablation}  
  \centering 
\begin{threeparttable} 
\begin{tabular}{l rrrr} 
        \toprule[0.4mm]
    Program &\method-NQ-NS &\method-NS & \method-AllS & \method \\ 
    \noalign{\smallskip}\hline
    \noalign{\smallskip}
     objdump  &9133.00& 9926.17  &  9752.33& \textbf{10287.67}\\
      libjpeg   &3816.67    & 3829.00    & 3839.33 & \textbf{3847.50} \\ 
        tcpdump   &14336.67 & 16140.67  & 14102.50 & \textbf{16549.67}\\
        libarchive  &7916.83  & 8060.50 & 7969.33 &\textbf{8116.83} \\
        pngimage    & 2413.17  & 2447.33  & 2408.50  & \textbf{2472.00} \\
        jhead    &\textbf{ 1091.83 }& 1088.83   &1082.83 & 1089.17 \\
        pngfix   &  2778.17  & 2794.17  & 2816.50  & \textbf{2835.33}\\
        libxml2  & 7430.00 &  7614.00   &8504.67& \textbf{9052.50}\\
        readelf   & 8977.00  & 9619.33  & 9401.50 & \textbf{9848.50}\\
        openjpeg  & 11238.17& 11836.67  & 11816.67 & \textbf{11872.17} \\
        gdk    &2432.17   & 2443.17  &\textbf{2502.17}& 2499.67\\
        nm     &5640.50     & 5837.50  & 5733.67 &\textbf{ 5918.00 }\\
        strip     &  8693.67        &  9365.17        & 9533.83 &  \textbf{9841.50}\\              
        imginfo    & 4780.33   & 5199.50   & 5099.83  & \textbf{5983.33}\\
        cyclonedds  & 4470.50 & 4466.17  &4474.17& \textbf{4498.83}\\
        \hline \noalign{\smallskip}
      \makecell[l]{\scriptsize{average} \\[-1pt] \scriptsize{improvement}} & &+4.03\%  & +3.47\% & +7.97\% \\
       \hline \noalign{\smallskip}
        \makecell[l]{\scriptsize{average} \scriptsize{improvement}\\[-1pt] \scriptsize{($>$10KLOC)}} & &+5.05\%  & +4.16\%& +9.70\%\\

        \bottomrule[0.4mm]
    \end{tabular}
    
    \end{threeparttable} 
\end{table}

Our method consists of three components progressively added to existing hybrid testing tools: (1) designing a new architecture to prevent over-pruning branches and support advanced coordination mechanism, (2) adopting a search strategy to explore high-value branches first, and (3) sampling at hard and high-reward branches. 
In the ablation study, we integrate these three components incrementally and evaluate the following three configurations.
\begin{itemize}
    \item \textbf{\method-NQ-NS} uses \method's architecture to solve branches that otherwise would be pruned by LOB strategy, but disables both the search strategy and sampling. The symbolic executor explores seeds in the order in which the coordinator synchronizes them into the execution tree.
    This configuration is used as the baseline of the ablation study.

    \item \textbf{\method-NS} retains \method's architecture and search strategy but disables sampling, aiming to assess the effectiveness of the search strategy.
    
    \item \textbf{\method-AllS} keeps \method's components except for sampling at all branches explored by symbolic execution, aiming to evaluate the sampling strategy (\textit{i.e.}, where to sample).
\end{itemize}
Table \ref{tab:ablation} shows experimental results of the ablation study.

\textbf{\method-NQ-NS}.
Across the seven programs on which SymCC spends more than 75\% of its time on sleeping (\emph{libarchive}, \emph{libpng}, \emph{jhead}, \emph{pngfix}, \emph{nm}, \emph{gdk}, and \emph{cyclonedds}), \method-NQ-NS improves coverage by an average of 3.7\% over SymCC. This result indicates that the branches pruned by LOB strategy are worth exploring. However, when considering all benchmark programs, the average coverage of \method-NQ-NS is slightly lower than that of SymCC. This demonstrates the necessity of a good search strategy.

\textbf{\method-NS}. As shown in Table \ref{tab:ablation} (column 3), \method-NS outperforms \method-NQ-NS by an average of 4.03\% across all programs and by 5.05\% on the 12 programs exceeding 10K LOC, demonstrating that the search strategy effectively improves testing efficiency.

\textbf{\method-AllS}. \method-AllS samples all branches explored by symbolic execution, like CoFuzz and PANGOLIN. This configuration is better than \method-NQ-NS, but worse than \method-NS. Thus, sampling everywhere does not necessarily bring benefit. Furthermore, the full version \method\ achieves 4.31\% higher coverage than \method-AllS, demonstrating the effectiveness of the sampling strategy.

In summary, we can obtain the following conclusion.
\begin{mdframed}[
skipabove=2pt,
    innertopmargin=2pt,
    innerbottommargin=2pt,
    linewidth=0.4pt]
	\textbf{Answer to Q3}: Sampling should be performed at hard branches with high expected reward. Our sampling strategy effectively improves testing efficiency.
\end{mdframed}


\begin{table*}
    \centering
    \begin{threeparttable}
    \caption{Bugs detected by \method\ in all experiments but not found by other tools.}
    \label{tab:bugs detected}
    
    \begin{tabular}{c@{} c@{} c@{} c c l c c c c}
    \toprule[0.4mm]
    Program & No. & Vulnerability & Bug Type & Bug Status & Tool &
    \scriptsize{RB} & \scriptsize{RB-ancestor} &
    \scriptsize{Sampling} & \scriptsize{Sampling-ancestor} \\
    \hline
    
    \multirow{1}{*}{readelf} 
     & 1 & \scriptsize{dum\_ia64\_vms\_dynamic\_fixups} & invalid memory read & Patched
     & \method & \redcheck & \redcheck & & \\
    \arrayrulecolor{gray!70}\hline
    
    \multirow{2}{*}{imginfo}
     & 2 &  \scriptsize{CVE-2017-5505} & invalid memory read & Patched
     & \method, \method-NS & & \redcheck & & \\
     & 3 & \scriptsize{CVE-2017-5504}& invalid memory read & Patched
     & \method & & \redcheck & & \\
    \arrayrulecolor{gray!70}\hline
    
    \multirow{5}{*}{cyclonedds}
     & 4 & idl\_vsnprintf & stack-overflow & Confirmed & \method & & & & \\
     & 5 & idl\_construct & invalid memory read & Patched & \method & & & & \\
     & 6 & idl\_log & stack-overflow & Patched & \method-NS & & & & \\
     & 7 & strlcpy & stack-buffer-overflow & Confirmed & \method 
         & \redcheck & & \redcheck & \\
     & 8 & do\_msg & invalid memory read & Confirmed & \method & & & & \\
    \arrayrulecolor{gray!70}\hline
    
    \multirow{8}{*}{objdump}
     & 9 & cache\_bread\_1 & heap-buffer-overflow & Patched
         & \method, \method-NS & \redcheck & \redcheck & & \\
     & 10 & \scriptsize{CVE-2019-9074} & invalid memory read & Patched
         & \method & & & \redcheck & \\
     & 11 & \scriptsize{CVE-2018-13033} & allocation-size-too-big & Patched
         & \method, \method-NS & \redcheck & \redcheck & \redcheck & \\
     & 12 &  \scriptsize{CVE-2017-14939} & heap-buffer-overflow & Patched
         & \method, \method-NS & \redcheck & & & \\
     & 13 & \scriptsize{CVE-2017-14129} & heap-buffer-overflow & Patched
         & \method & & \redcheck & \redcheck & \\
     & 14 & \scriptsize{CVE-2017-15939} & invalid memory read & Patched
         & \method & \redcheck & \redcheck & & \\
     & 15 & bfd\_get\_section\_contents & negative-size-param & Patched
         & \method & & & & \\
     & 16 & \scriptsize{\_bfd\_elf\_canonicalize\_dynamic\_reloc}
           & heap-buffer-overflow & Patched
           & \method & & \redcheck & & \redcheck \\
    \arrayrulecolor{gray!70}\hline
    
    \multirow{1}{*}{strip}
     & 17 & \scriptsize{CVE-2017-14529} & heap-buffer-overflow & Patched
     & \method, \method-NS & & \redcheck & & \\
    \arrayrulecolor{black}
    \bottomrule[0.4mm]
    
    \end{tabular}
    \end{threeparttable}
    \end{table*}

\subsection{\textbf{Crash Analysis}}

To investigate how our approach contributes to crash discovery, we analyze the crashes listed in Table \ref{tab:bugs detected}, which are detected by \method\ but missed by other tools. 
We mark branches pruned by the LOB strategy but solved in \method\ as \textit{Retrieved Branch} (RB). According to the relation between crash and retrieved branch or sampling, we mark these crashes using the following four labels (Columns~7--10 in Table \ref{tab:bugs detected}).

\begin{enumerate}
    \item  $\mathsf{RB}$: The crash is directly triggered by a seed generated from solving a retrieved branch.
    \item \textbf{$\mathsf{RB}$-ancestor:} The crash is triggered by a seed that has at least one ancestor generated from solving a retrieved branch. In hybrid testing, a seed $s$ may be iteratively mutated by fuzzing or changed by solving open branches in symbolic execution. In such a case, seed $s$ is called the ancestor of its propagated seeds.  
    \item $\mathsf{Sampling}$: The crash is directly triggered by a seed obtained through sampling.
    \item \textbf{$\mathsf{Sampling}$-ancestor:} The crash is triggered by a seed that has at least one ancestor generated through sampling.
\end{enumerate}

The first two labels are RB-related, and the last two are sample-related. 
In the experiments, a single crash may be tagged by multiple labels when it can be triggered by multiple seeds. 

As shown in Table \ref{tab:bugs detected}, among the 17 crashes, 6 are tagged as $\mathsf{RB}$. 
9 crashes are tagged as $\mathsf{RB}$-ancestor. 
In total, 11 crashes are related to $\mathsf{RB}$. 
These results demonstrate the benefit of retaining branches that would otherwise be pruned by LOB strategy. 

Additionally, 4 crashes are tagged with $\mathsf{Sampling}$. Among these 4 crashes, one crash (\texttt{bfd\_get132} in \emph{objdump}) can be only triggered by $\mathsf{Sampling}$. Another one crash is tagged with $\mathsf{Sampling}$-ancestor. These results demonstrate that sampling can help discover crashes.

In total, 12 out of 17 crashes are explicitly related to $\mathsf{RB}$ or $\mathsf{Sampling}$. Importantly, 11 out of these 12 crashes cannot be triggered without $\mathsf{RB}$ or $\mathsf{Sampling}$. The remaining 5 crashes, which are not explicitly related to our method, but still may benefit from the search strategy. These results demonstrate the effectiveness of our method in discovering crashes.

In the following, we analyze one RB-related and one sampling-related crash.

\subsubsection{\textbf{Crash Directly Triggered by Solving RB Branch}}

\begin{figure}[h]
	\setlength{\intextsep}{0pt}  
	\setlength{\textfloatsep}{0pt} 
	\setlength{\floatsep}{0pt}   
	\setlength{\abovecaptionskip}{-1pt} 
	\setlength{\belowcaptionskip}{1pt} 
	\centering
	\hfill 
	\begin{lstlisting}[basicstyle=\fontsize{8pt}{8pt}\ttfamily, lineskip={0pt}]
file_ptr cache_bread (void *buf, file_ptr nbytes)
{//nbytes -> size of .debug_info
    const file_ptr max_chunk_size = 0x800000;
    file_ptr chunk_size = nbytes - nread;
    if (chunk_size > max_chunk_size)
        chunk_size = max_chunk_size;
    cache_bread_1((char*) buf + nread,chunk_size);
}
file_ptr cache_bread_1 (void *buf, file_ptr nbytes)
{ 
    nread = fread(buf,1,nbytes,f);//heap overflow
}  
	\end{lstlisting}   
	\caption{Crash in \emph{objdump}'s \texttt{cache\_bread\_1} function.}
	\label{fig:src1}
\end{figure}
\emph{objdump} is a powerful command-line utility for displaying various information about object files on Unix-like operating systems.
The heap overflow vulnerability (No. 9 in Table~\ref{tab:bugs detected}) in the \texttt{cache\_bread\_1} function of \emph{objdump} (see Fig. \ref{fig:src1}) is directly triggered by solving a retrieved branch.  
Function \texttt{cache\_bread} has two parameters \texttt{buf} and \texttt{nbytes}. The size of \texttt{buf} and the value of \texttt{nbytes} are both parsed from the input binary.
The crash is caused by an excessively large  \texttt{nbytes} (\texttt{0xFFFFBFFFFFFFE00}), which is parsed from the \texttt{.debug\_info} section of the input binary. Although line 5 clips the read size to 8 MB, the buffer \texttt{buf} has a capacity of only 465 bytes. Consequently, invoking \texttt{fread} (line 11) with 8 MB as \texttt{nbytes} triggered a heap overflow.

During testing, a seed containing an abnormal \texttt{nbytes} in \texttt{.debug\_info} section is generated. But the seed does not trigger the heap overflow. When the seed is executed by symbolic execution, a $\mathsf{RB}$ branch is solved, and the heap overflow is triggered on the path of the resulting solution. Other fuzzing tools and hybrid testing tools that rely on the LOB strategy fail to reveal this vulnerability.

\begin{figure}[h]
    \setlength{\intextsep}{-2pt}  
    \setlength{\textfloatsep}{0pt}
    \setlength{\floatsep}{0pt}    
    \setlength{\abovecaptionskip}{-3pt} 
    \setlength{\belowcaptionskip}{1pt} 
    \centering
      \begin{lstlisting}[basicstyle=\fontsize{8pt}{8pt}\ttfamily, lineskip={0pt}] 
bfd_byte *info_ptr = stash->info_ptr;        
bfd_byte *end_ptr  = info_ptr + unit_length; 
info_ptr += blk->size;    
offset = read_4_bytes (info_ptr, end_ptr);//dwarf2.c
int read_4_bytes(bfd_byte *buf, bfd_byte *end){
    if (buf + 4 > end)       
       return 0;    
    return bfd_getl32(buf);
}
bfd_vma bfd_getl32(const void *p){   
    const bfd_byte *addr = (const bfd_byte *) p;  
    unsigned long v=(unsigned long)addr[0];//crash
  \end{lstlisting}
    \caption{Crash in objdump's \texttt{bfd\_getl32} usage.}
    \label{fig:src2}
  \end{figure}

\subsubsection{\textbf{Crash Directly Triggered by Sampling}}

\begin{table}[t]
	
	\caption{Average edge coverage of \method\ for 3 runs of 24 hours with different $\delta$ and $\gamma$ settings. For the best configuration, we use the average results of 6 times, as in the main results.}
	\label{tab:DR}       
		\begin{tabular}{l C{0.6cm} C{0.6cm} C{0.6cm} C{0.9cm} C{0.6cm} C{0.6cm} C{0.6cm}}
			
					\toprule[0.4mm]
					$\log \delta$ & $0$ & $0$ & $-150$ & $-150$ & $-150$ & $-300$  & $-300$ \\
					\hline
					$\gamma$ & $0$ & $300$ & $0$ & $300$ & $600$ & $300$  & $600$ \\
					\hline
					objdump & 9855 & 10025 & 10089  & 10288  & 9856  & \textbf{10858 }  & 10222  \\
					libjpeg & 3827 & 3848  & \textbf{3881}  & 3848  & 3836  & 3845  & 3829 \\
					tcpdump & 15605 & 15738 & 15747 & \textbf{16550} & 16233 & 16206  & 16301 \\
					libarchive & 8095 & 8019 & 8078  & \textbf{8117} & 8080 & 8114  & 8021 \\       
					pngimage & 2451 & 2424  & 2429   & \textbf{2472 } & 2452  & 2453   & 2411  \\
					jhead & 1092 & 1101  & 1094 & 1089  & 1096  & \textbf{1102 }  & 1101 \\
                    pngfix & 2827 & 2758  & 2818 & 2835  & 2751  & 2781   & \textbf{2839 } \\ 
					libxml2 & 7045 & 8486 &  7051  & \textbf{9053}  & 7906  & 8225   & 7647  \\					 
					readelf & 9221 & 9389 & 9848  &  \textbf{9849 } & 9739  & 9579   & 9319  \\
                    openjpeg & \textbf{11920}  & 11837 & 11672  & 11872  & 11801  & 11732   & 11745 \\
					gdk & 2471 & 2496 & 2414 & 2500 & \textbf{2534 } & 2506   & 2449  \\
                    nm & 6092 & 5284  & 5697 & 5918  & 5906  & 6135   & \textbf{6156 } \\
					strip &8994 &8798 & 9362 &\textbf{9842}& 8894 &9586 & 8841\\
					imginfo & 5577 & 4763  & 5133 & \textbf{5983 } & 5389  & 5837   & 5100  \\
                    cyclonedds &  4469 & 4479  & 4464 & 4499 & 4467  & 4469   & \textbf{4505 } \\
					\hline
					\noalign{\smallskip}
					average & 6636 &6630  &6652    &\textbf{6981}   & 6736  &6895   &6699  \\
					\bottomrule[0.4mm]
				\end{tabular}
			\end{table} 

  The invalid memory read (No.~10 in Table~VIII) in the \texttt{bfd\_getl32} function of \texttt{objdump} (see Fig. \ref{fig:src2}) is triggered by sampling.
  \texttt{info\_ptr} points to the \texttt{.debug\_info} buffer with 58 bytes, which is parsed from the input binary file.
  \texttt{end\_ptr} is assigned as \texttt{info\_ptr + unit\_length}.
  An abnormal \texttt{unit\_length} value (1,740,402,063) parsed from the input causes \texttt{end\_ptr} to exceed the buffer boundary.
  Similarly, an abnormal \texttt{blk->size} value (1,682,180,503) causes \texttt{info\_ptr} (and thus \texttt{buf}) to advance beyond the end of the buffer.
  Because the boundary check relies on this incorrect \texttt{end} pointer, the out-of-bounds \texttt{buf} is not detected and is dereferenced, resulting in an invalid memory read.

 During testing, a seed containing normal values (17 and 0) in the \texttt{.debug\_info} section is generated first. Then, a seed with anomalous values (\texttt{1,740,402,063} and \texttt{1,682,180,503}) is generated during $\mathsf{Sampling}$, causing both \texttt{end} and \texttt{buf} to exceed the actual buffer bounds and trigger the crash.

\subsection{Parameter Configuration}
In Algorithm \ref{alg:searchAndSampleStrategy}, we use two thresholds, the fuzzing difficulty threshold $\delta$ and future reward threshold $\gamma$, to control whether to perform sampling (line \ref{alg:solveorsample_start}), and another parameter $\lambda$ (line \ref{alg:score}) to control the priority of open branches. It is too costly to find the best configuration due to the explosion of parameter compositions. 
Since the two thresholds $\delta$ and $\gamma$ affect sampling chances, and $\lambda$ affects branch priority. We treat $\delta, \gamma$ and $\lambda$ as two separate groups when seeking the best parameter.

  \subsubsection{\textbf{Impact of $\delta$ and $\gamma$}}
We perform a grid search to investigate the impact of $\delta$ and $\gamma$ on edge coverage. After observing the ranges of fuzzing difficulty and future reward in practice, we set the range of $\log \delta$ as $[-300,0]$ with a step of $-150$, the range of $\gamma$ as $[0, 600]$ with a step of 300. For example, setting $\delta=1\ (\textit{i.e.}, \log \delta=0)$ and $\gamma=0$ is the most relaxed configuration, which specifies sampling at every branch with non-zero future reward.

Table \ref{tab:DR} shows the results of using different thresholds. 
Overall, the combination of $\log \delta=-150$ and $\gamma=300$ performs the best, followed by $\log \delta=-300$ and $\gamma=300$. 
These results accord with our expectations.
When the thresholds are too relaxed, the system samples branches that are neither difficult to fuzz nor expected to provide high future reward, resulting in redundant test inputs. 
In contrast, overly strict thresholds reduce sampling opportunities and cannot utilize the power of sampling. Therefore, a moderate threshold performs the best. We omit other edge configurations in Table \ref{tab:DR} because they perform worse than the best one.

\subsubsection{\textbf{Impact of $\lambda$}}
\begin{table}[t]
    \setlength{\abovecaptionskip}{2pt} 
\caption{Average edge coverage of  \method\ for 3 runs of 24 hours with different $ \lambda $ settings. For the best configuration, we use the average results of 6 times as in the main results.}
\centering 
\label{tab:lamda}       
       \begin{tabular}{l rrrr}
           \toprule[0.4mm]
           \noalign{\smallskip}
           Program     & $\lambda=0.1$(used) & $\lambda=0.4$ & $\lambda=0.7 $ 
           \\\noalign{\smallskip}\hline\noalign{\smallskip}
           objdump      &\textbf{10287.67} & 9917.33 & 9914.67 \\
           libjpeg      & 3847.50 & \textbf{3848.67} & 3828.00 \\
           tcpdump      & \textbf{16549.67} & 15756.67 & 15202.67  \\
           libarchive    & \textbf{8116.83} &7991.33 & 8074.00  \\
           pngimage      & 2472.00 & 2444.33& \textbf{2493.33} \\
           jhead         &  1089.17 &1094.67  &\textbf{1097.67}\\
           pngfix       & 2835.33 & \textbf{2837.67} & 2826.33 \\
           libxml2       & 9052.50  & 8567.00 & \textbf{9353.00} \\    
           readelf      &  9848.50 & 9771.33 &  \textbf{9891.67} \\
             openjpeg      & 11872.17 & \textbf{11969.33} & 11819.67  \\
            gdk           & 2499.67 & \textbf{2502.00} & 2475.67\\
           nm           &  \textbf{5918.00} & 5604.67 & 5371.33  \\
           strip         & 9841.50  & 9585.00   &\textbf{ 10586.00} \\
           imginfo      & \textbf{5983.33} & 5232.00 & 5284.00 \\
           cyclonedds   &\textbf{ 4498.83} & 4473.67 & 4466.33\\
            \hline
            \noalign{\smallskip}
           average & \textbf{6980.84} & 6773.04& 6845.62  \\
           \bottomrule[0.4mm]
       \end{tabular}
\end{table} 
Table \ref{tab:lamda} presents the impact of $\lambda$, which controls the balance between fuzzing difficulty and future reward in calculating branch priority. We set $\lambda$ as 0.1, 0.4, and 0.7 for comparison. The best result in each row is in bold. When $\lambda$ takes 0.1, \method\ achieves the best performance.

\subsection{{\method\ With AFL++}}
In the above experiments, all hybrid testing tools integrate AFL as the fuzzer. We replace AFL with AFL++, a more advanced fuzzer, in \method\ to show that our method can be improved by integrating more advanced fuzzing techniques.
We note \method\ with AFL++ as \method-AFL++. 
As shown in Table \ref{tab:s2fafl++}, \method-AFL++ achieves higher edge coverage than \method, demonstrating that our method benefits from advanced fuzzers.

\begin{table}[t]
	\setlength{\abovecaptionskip}{2pt} 
	\caption{Average edge coverage of  \method-AFL++ for 6 runs of 24 hours.}
	\centering 
	\label{tab:s2fafl++}      
	\begin{tabular}{l rrrr}
		\toprule[0.4mm]
		\noalign{\smallskip}
		Program    &AFL & \texttt{\method}(AFL)  & AFL++  & \texttt{\method-AFL++} 
		\\\noalign{\smallskip}\hline\noalign{\smallskip}
		objdump     &8652.67&10287.67 &9380.33 & 10470.17 \\
		libjpeg     &3825.00 &3847.50& 3818.00 & 3824.16  \\
		tcpdump     & 81.00&16549.67 &81.00 & 16074.00   \\
		libarchive   &5139.50 &8116.83 & 5206.17 &8108.50   \\
		pngimage     &1671.83 &2472.00&1660.17 & 2575.50 \\
		jhead        &388.00 &1089.17 & 387.83 &1101.83 \\
        pngfix       &1816.00&2835.33& 1765.83 &2800.17\\  
		libxml2      &6971.33 &9052.50&7951.00  & 9960.83 \\
		readelf      &8873.67 &9848.50 &9393.83 & 9799.00  \\
        openjpeg   &11364.50  &11872.17 & 12212.33 & 12307.00   \\
        gdk          &2449.50&2499.67 &2481.33 & 2492.33\\
		nm          &4646.00&5918.00 &  5223.67 & 6766.50  \\
		strip       &7943.00 &9841.50 & 8229.67  & 9606.00 \\
		imginfo    &3970.33 &5983.33 & 4461.00  & 5810.83 \\
        cyclonedds  &4453.50&4498.83 &4491.33 & 4526.50\\
		\hline
		\noalign{\smallskip}
		average &4816.41&6980.84& 5116.23 & \textbf{7081.55}  \\
		\bottomrule[0.4mm]
	\end{tabular}
\end{table}

\section{Discussion and Future Work} \label{sec:Discussion}

\textit{\textbf{Architecture.}}
The architecture of \method\ provides a flexible infrastructure that can integrate various components in hybrid testing, including advanced search strategies, seed scheduling algorithms, and other coordination mechanisms. 
For example, the principles proposed in this paper can be implemented in a different way under the current architecture. 
In the future, we will explore other algorithms to estimate the fuzzing difficulty and reward of open branches.

\textit{\textbf{Sampling.}}
\method\ is the first work to systematically incorporate fuzzing, constraint solving, and sampling. Our work demonstrates that sampling should be performed on appropriate branches.  
Many current sampling algorithms aim to generate samples uniformly distributed in the solution space rather than in the path space.
In the future, we plan to explore sampling methods more suitable for software testing.



\section{Related Work} \label{sec:relatedwork}

\textit{\textbf{Symbolic Execution}.}
KLEE \cite{KLEE} enables comprehensive path exploration using fine-grained search strategies. SymCC\cite{SymCC} instruments programs at the LLVM IR level during compilation, which significantly reduces the overhead of symbolic computation. 
SYMSAN \cite{SymSAN} reduces the cost of managing symbolic expressions by integrating symbolic execution with LLVM-based data-flow analysis. SymQEMU \cite{SymQemu} extends the applicability of SymCC to binary programs by leveraging QEMU-based dynamic binary translation. Marco \cite{Marco} models the path divergence rate at each symbolic branch, treats concolic execution as a Markov process, and prioritizes paths with smaller divergence rates. Marco focuses on symbolic execution rather than a hybrid testing tool, and it does not employ sampling.
The above approaches are orthogonal to \method.

\textit{\textbf{Hybrid Testing}.}
In \cite{hybrid2007}, Sen et al. propose combining random testing with concolic execution for testing reactive systems. Subsequently, Driller~\cite{Driller} is among the first to integrate coverage-guided fuzzing with symbolic execution.
Angora~\cite{Angora} increases branch coverage by solving path constraints without symbolic execution, leveraging byte-level taint tracking and a gradient-descent–based search for constraint solving.
QSYM \cite{QSYM} integrates symbolic emulation with native execution through dynamic binary translation and introduces pruning strategies to improve the efficiency of hybrid testing. 
\cite{rare} identifies rare paths in programs and guides symbolic execution to explore these paths. It does not execute fuzzing and symbolic execution in parallel.

Seed scheduling is a key research problem in hybrid testing. QSYM, SymCC, SYMSAN, and SymQEMU \cite{QSYM, SymCC, SymSAN, SymQemu} employ simple seed features to determine seed priority. DigFuzz \cite{Dig_Fuzz} leverages the Monte Carlo method to estimate path difficulty and sends the hardest paths for symbolic execution. 
 MEUZZ \cite{MEUZZ} uses a linear model to determine which seeds are sent first to symbolic execution. The parameters of the linear model are adjusted during the testing process. These tools run fuzzing and symbolic execution in parallel and are built on tailored symbolic executors with the LOB strategy. Therefore, these approaches cannot handle the sleep issue.

\textit{\textbf{Sampling}.}
Legion \cite{Legion} introduces SMT sampling into symbolic execution and uses the Monte Carlo tree search to develop an optimal search strategy. Since Legion is not a hybrid testing tool, we did not compare it.
PANGOLIN \cite{Pangolin} applies symbolic execution to hard branches, utilizes the BWAI \cite{BWAI} for path abstraction, and samples within the polyhedron via the Dikin walk \cite{Dinkin}. CoFuzz \cite{Co_Fuzz} uses a linear model to schedule the priority of branches for symbolic execution, utilizes the interval path abstraction and John walk \cite{Jhon_Walk} method sampling. PANGOLIN \cite{Pangolin} and CoFuzz \cite{Co_Fuzz} also use tailored symbolic executors and LOB strategy. 
PANGOLIN \cite{Pangolin} and CoFuzz \cite{Co_Fuzz} sample all branches explored by symbolic execution, whereas \method\ samples only appropriate branches and solves the rest.

To the best of our knowledge, \method\ is the first work to systematically explore the integration of fuzzing, symbolic execution, and sampling.

\section{Conclusion}\label{sec:conclude}
In this paper, we propose a novel hybrid testing tool \method\ integrating fuzzing, symbolic execution, and sampling, aiming to address the sleeping issue and sampling strategy in hybrid testing. Our method consists of a new hybrid testing architecture that combines the strengths of conventional and tailored symbolic executors. Based on this architecture, we propose principles for integrating different techniques. Experimental results show that our method outperforms the SOTA hybrid testing tool by 6.14\% in edge coverage and 32.6\% in discovered crashes. %

\textbf{Artifacts}. The artifacts related to this work can be found at \url{https://github.com/research-anonymous-se/S2F-artifacts}.

\bibliographystyle{IEEEtran}
\bibliography{sample-base}

\end{document}